\documentclass[journal,twocolomns]{IEEEtran}
\usepackage{amsmath}
\usepackage{graphicx}
\usepackage{amssymb,amsmath}
\usepackage{mathtools}
\usepackage{hyperref}
\usepackage{cleveref}
\usepackage{autonum}
\usepackage[usestackEOL]{stackengine}
\usepackage[normalem]{ulem}
\usepackage{balance}

\usepackage{subfigure}
\usepackage{graphicx,epsfig}
\usepackage[linesnumbered,ruled,vlined]{algorithm2e} 
\usepackage{empheq}
\usepackage{cite}
\usepackage{color}
\usepackage[normalem]{ulem}

\newtheorem{theorem}{Theorem}

\newcommand{\mt}[1]{}
\let \mt=\mathrm

\makeatletter

\newcommand{\Rmnum}[1]{\expandafter\@slowromancap\romannumeral #1@}

\makeatletter

\newcommand{\comm}[1]{}

\let \bd = \textbf
\let \bs = \boldsymbol

\def \p1#1{#1^{-1}}

\def \~#1{\tilde{#1}}

\ifCLASSINFOpdf
\else
\fi
\begin{document}
%
\title{Switch-based Hybrid Beamforming for Massive MIMO Communications in mmWave Bands}

\author{\IEEEauthorblockN{Hamed Nosrati\IEEEauthorrefmark{1}\IEEEauthorrefmark{3}, Elias Aboutanios \IEEEauthorrefmark{1}\IEEEauthorrefmark{3}, Xiangrong Wang \IEEEauthorrefmark{2}, and
David Smith\IEEEauthorrefmark{3}\IEEEauthorrefmark{4}}\\
\IEEEauthorblockA{\IEEEauthorrefmark{1}School of Electrical and Telecommunications Engineering,
University of New South Wales Australia\\}
\IEEEauthorblockA{\IEEEauthorrefmark{2} School of Electronic and Information Engineering
Beihang University, Beijing, China}\\
\IEEEauthorblockA{\IEEEauthorrefmark{3} Data61, CSIRO Australia}, \IEEEauthorblockA{\IEEEauthorrefmark{4} Australian National University}\\
\IEEEauthorblockA{hamed.nosrati@unsw.edu.au,
elias@unsw.edu.au,
xrwang@buaa.edu.cn, david.smith@data61.csiro.au}
}


%


\maketitle
\begin{abstract}
Switch-based hybrid network is a promising implementation for beamforming in  large-scale millimetre wave (mmWave) antenna arrays. By fully exploiting the sparse nature of the mmWave channel, such hybrid beamforming reduces complexity and power consumption when compared with a structure based on phase shifters. However, the  difficulty of designing an optimum beamformer in the analog domain is prohibitive due to the binary nature of such a switch-based structure. Thus, here we propose a new method for designing a switch-based hybrid beamformer for massive MIMO communications in mmWave bands. We first propose a method for decoupling the joint optimization of analog and digital beamformers by confining the problem to a rank-constrained subspace.  
 We then approximate  the solution through two approaches: norm maximization~(SHD-NM), and majorization~(SHD-QRQU). In the norm maximization method, we propose a modified sequential convex programming (SCP) procedure that maximizes the mutual information while addressing the mismatch incurred from approximating the log-determinant by a Frobenius norm. In the second method, we employ a lower bound on the mutual information by QR factorization. We also introduce linear constraints in order to include frequently-used partially-connected structures. Finally, we show the feasibility, and effectiveness of the proposed methods through several numerical examples. The results demonstrate ability of the proposed methods to track closely the spectral efficiency provided by unconstrained optimal beamformer and phase shifting hybrid beamformer, and outperform a competitor switch-based hybrid beamformer.  
 
\end{abstract}




%
\IEEEpeerreviewmaketitle
\section{Introduction}\label{intro}

\IEEEPARstart{M}assive multiple-input multiple-output (MIMO) systems in millimetre wave (mmWave) bands are promising candidates for future generation wireless cellular communications systems to alleviate spectrum congestion and bandwidth scarcity~\cite{Han2015}. Communications in mmWave band, which makes use of frequency bands from 30 to 300 GHz, is an enabling technology for fifth-generation (5G) networks. The smaller wavelengths in mmWave communications systems make \mbox{large-scale} antenna arrays at the transceivers viable. This leads to various new challenges for fully-digital beamforming in mmWave massive MIMO systems, such as prohibitively high hardware complexity, computational cost and power consumption~\cite{Zhang2015,Heath2016,Ayach2014}.

Digital processing for traditional MIMO communications systems requires that each array element has a dedicated RF and baseband hardware chains comprising expensive components~\cite{Rusu2016}. Thus, full digital processing is undesirable and impractical due to the cost and complexity of the hardware chains in mmWave MIMO communications systems with large arrays~\cite{Sohrabi2015,Bogale2014}. Furthermore, components such as RF up/down converters and analog-to-digital converters (ADCs) (or digital-to-analog converters (DACs)) are not only expensive but also have high power consumption~\cite{Alkhateeb2014}. This motivates various strategies for the efficient implementation of beamforming for massive MIMO systems in mmWave bands, including hybrid beamforming architectures~\cite{Nsenga2010,Roh2014,Gholam2011,Pi2012,Zhang2005,Venkateswaran2010,Sayeed2010,Ayach2014,Alkhateeb2014,Han2015}, beamspace signal processing techniques~\cite{Brady2013,Sayeed2002}, lens-based analog beamforming antennas~\cite{Gao2018}, and low-rate ADC methods~\cite{Mo2014}. 

Hybrid beamforming for massive MIMO communications systems has been extensively studied in recent years, see~\cite{Molisch2017} and the references therein. It is a well-established approach that employs a two-stage analog and digital processing configuration. The analog precoding stage employs simpler and less power-hungry analog beamformers to present a reduced-dimensional signal to the baseband stage~\cite{Jiang2018}. Typically, the analog precoding stage is implemented as a network of phase shifters~\cite{Mendez-Rial2016}. While this achieves a level of simplification over fully digital RF chains, the practical realization of the phase shifters for mmWave frequencies is not a simple task~\cite{Poon2012}. MmWave digitally controlled phase shifters have finite precision that may not be sufficient to form the desired beams~\cite{Poon2012}, and their latency may lead to performance degradation if the channel is rapidly changing~\cite{Molisch2017}. Their phase precision can be improved but at the cost of higher power consumption~\cite{Ahmed2018}. Passive phase shifters, on the other hand, can alleviate the power consumption problem, but they are known to incur higher losses, requiring additional amplification to maintain an acceptable output signal-to-noise ratio (SNR)~\cite{Poon2012}. 

The challenges associated with the use of phase shifters have motivated research into alternative approaches for the realization of the analog precoder. In particular, switch-based networks are simple, low-power and high-speed solutions to these challenges~\cite{Molisch2017,Mendez-Rial2016}. These switch-based combiners effectively combine subsets of the available antennas such that they are able to leverage the sparse nature of mmWave massive MIMO channels to realize performance gains~\cite{Ahmed2018}. In fact, different realizations and special cases of the switch-based approach have been successfully applied in various contexts. For instance, optimal selection of a subset of ``best'' antennas from a larger set of antennas has been shown as a promising approach for delivering a large aperture with satisfactory performance at reduced hardware cost and complexity~\cite{Molisch2004,Gharavi-Alkhansari2004,Nosrati2017,Wang2014,Nosrati2017a,Amin2016}. When the complexity of connectivity, routing, and RF multiplexing are of grave concern, the switching network can be partitioned into subsets, and only a few antennas per subset are then selected~\cite{Mendez-Rial2016,Ayach2013,Nosrati2018}.

The solution to the hybrid beamforming problem is not straightforward, and replacing the phase shifters with simple switches greatly exacerbates the difficulty of this task as optimization over a set of binary variables is then required.
A dictionary-based strategy was proposed in~\cite{Mendez-Rial2015} to address this, but the dictionary grows exponentially with the number of antennas, making this method impractical for large antenna arrays. Furthermore, as this dictionary is scenario-specific, it must be redesigned for every problem variation, which adds another layer of complexity to it. In~\cite{Jiang2018} a unified greedy algorithm is proposed for the design of both phase-shifter and switch-based networks. However, the proposed greedy algorithm is limited to the case where the digital beamforming matrix is square with a dimension equal to the number of data streams to be transmitted. Moreover, the proposed greedy method is only suitable for the design of an unconstrained switch-based hybrid beamformer, as its formulation prevents the incorporation of any specific constraints.

Thus, here we propose a novel approach that has the capability of solving a switch-based hybrid beamformer design problem for massive MIMO communications in mmWave bands, which also has the flexibility to include a variety of desirable constraints. To this end, we formulate the problem as an optimization. We decouple the transmit and receive problems, and decompose the joint optimization of the analog and digital precoder matrices into a rank-constrained single variable problem. Then we proceed to solve this problem via two different strategies. In the first method, by approximating the mutual information via a Frobenius norm, we are able to cast the problem as a norm maximization. We then proceed to solve the non-convex maximization via a sequence of optimizations in a modified version of sequential convex programming (SCP). In order to address the mismatch between the norm maximization and the mutual information maximization, the SCP is guided by the actual value of mutual information in each step. In the second method, so as to reduce computational complexity, we take advantage of a lower bound given by QR factorization and iteratively optimize the columns of the analog precoder such that in each iteration we maximize a quadratic form via a SCP.

Hence, the contributions of this paper are: 
\begin{itemize}
\item We propose a method to solve the problem of switch-based hybrid beamforming for massive MIMO communications in mmWave bands based on convex optimization. This formulation allows for the examination of several cases with various practical limitations on hardware resources.
\item As the vast majority of hybrid design methods are based on the assumption that the optimal beamformer is realized by a combination of digital and analog beamformers, there is a mismatch between the approximated solution and the exact one. To address this issue, which is more extreme in a switch-based network due to the binary nature of the variables and the corresponding feasible solution space, we propose a heuristic method based on Gaussian randomization.
\item We propose two cost functions as surrogates for the maximization of the mutual information based on (1) a Frobenius norm approximation, and (2) a QR lower bound.
\item We study partially connected switch-based networks and propose a method to model specific requirements, imposed by an arbitrary partially connected network, using linear constraints.
\end{itemize}

\subsection{Organization}
The rest of this paper is organized as follows. We provide the system model and formulate the hybrid beamforming (precoding) in Section ~\ref{sec:problem_form}. In Section ~\ref{sec:proposal}, we present a new method for the design of switch-based hybrid beamforming. We propose two separate algorithms to solve the formulated problem in Section~\ref{sec:norm_max} and Section~\ref{sec:QRQU}, respectively.
We formulate the design problem of switch-based hybrid beamforming for partially connected networks in Section~\ref{sec:subsets}. Finally, in Section ~\ref{sec:sim}, we validate the effectiveness of the proposed method via numerical examples that are compared with existing state-of-the-art solutions.
\subsection{Notation}
In the remaining of the paper,  we use lower-case letters to denote
vectors, and upper-case letters for matrices. 
The notation $\mathbb E$ denotes the expectation operator, 
The notations $\mt{Tr}(\bd A)$, $\bd A^T$ and $\bd A^H$ denote trace, transposed and conjugate transpose of matrix $\bd A$. 
Matrix $\bd I_N$ is an identity matrix of size $N$, $\bd 1_N$ is a $N \times 1$ vector of all ones, and $\bd 0_N$ a vector of zeros. The operation $\|\bd A\|_ F$ denotes the Frobenius norm. We also use $\nabla$ to represent gradient. $\mt{vec}(\bd A)$ vectorizes the matrix $\bd A$ by stacking its columns. Moreover, the real part of $\bd A$ is shown by $\mt{real}(\bd A)$.
\section{Problem Formulation}\label{sec:problem_form}
\begin{figure*}[!t]
    \centering
    \includegraphics[width=17cm]{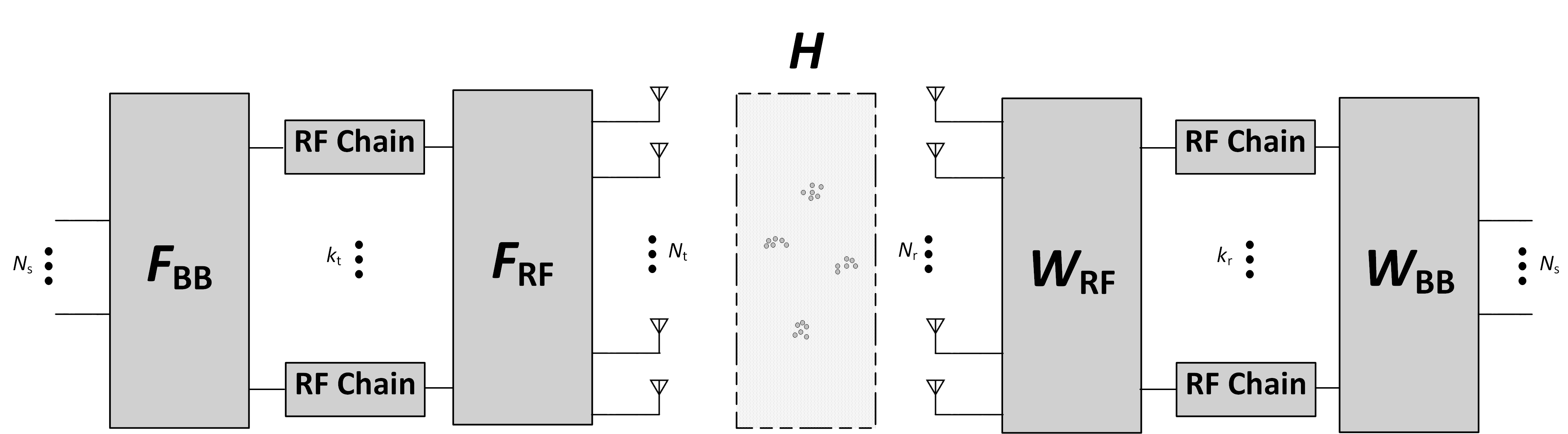}
    \caption{Block diagram of hybrid MIMO architecture for mmWave communication with baseband and analog precoder/combiner with a clustered channel model.}
    \label{fig:system-model}
\end{figure*}

A hybrid structure for a single-user mmWave MIMO system is depicted in Fig.~\ref{fig:system-model}. In this setup, the transmitter comprises $N_\mt{t}$ antennas and $k_\mt{t}$ RF transmit chains, and is required to send $N_\mt{s}$ data streams to the receiver. We assume that $ N_\mt{s}\leq k_\mt{t}\leq N_\mt{t}$. Let the transmit digital beamforming matrix be $\bd F_\mt{BB}$ of size $k_\mt{t}\times N_\mt{s}$, and RF precoder matrix be $\bd F_\mt{RF}$ of size $N_\mt{t}\times k_\mt{t}$. $\bd F_\mt{RF}$ is implemented using analog phase-shifters or RF switches. The discrete-time transmit signal is then $\bd x=\bd F \bd s$, where \mbox{$\bd F=\bd F_\mt{RF}\bd F_\mt{BB}$}, and $\bd s$ is the $N_\mt{s}\times 1$ symbol vector such that $\mathbb{E}[\bd s \bd s^H]=\frac{1}{N_\mt{s}}\bd I_{N_\mt{s}}$ with $\mathbb{E}$ denoting the expected value. At the receiver, $N_\mt{r}$ antennas are connected to $k_\mt{r}$ RF receive chains to recover the transmitted symbol $\bd s$. Similarly to the transmitter, the receive beamformer $\bd W=\bd W_\mt{RF}\bd W_\mt{BB}$ is composed of the $N_\mt{r}\times k_\mt{r}$ RF combining matrix $\bd W_\mt{RF}$ and $k_\mt{r}\times N_\mt{s}$ baseband beamforming matrix $\bd W_\mt{BB}$.
Given a narrowband frequency-flat channel model represented by the $N_\mt r \times N_\mt t$ channel matrix $\bd H$, with $\mathbb{E}\left[ \|\bd H\|_{F}^2 \right]=N_\mt{t}\times N_\mt{r}$, we can write the received signal as
\begin{align}
\nonumber \bd y=\sqrt{\rho}\bd W_\mt{BB}^H\bd W_\mt{RF}^H\bd H\bd F_\mt{RF}\bd F_\mt{BB}\bd s+\bd W_\mt{BB}^H\bd W_\mt{RF}^H\bd n.
\end{align}
Here $\rho$ is the average received power, and $\bd n $ the additive zero-mean i.i.d noise with variance $\sigma_n^2$. Also, $\bd W_\mt{BB}^H$ denotes the conjugate transpose of $\bd W_\mt{BB}$. For a clustered channel model consisting of the sum of the contributions of $N_\mt{cl}$ scattering clusters, with each cluster comprising $N_\mt{ray}$ propagation paths, the channel matrix is
\begin{align}
    \nonumber\bd H=\gamma\sum\limits_{i,\ell}\alpha_{i\ell}\Lambda_\mt r(\phi_{i\ell}^\mt r,\theta_{i\ell}^\mt r)\Lambda_\mt t(\phi_{i\ell}^\mt t,\theta_{i\ell}^\mt t)\bd a_\mt r(\phi_{i\ell}^\mt r,\theta_{i\ell}^\mt r)\bd a_\mt t(\phi_{i\ell}^\mt t,\theta_{i\ell}^\mt t)^H,
\end{align}
where $\gamma=\sqrt\frac{N_\mt{rt}N_\mt{r}}{N_\mt{cl}N_\mt{ray}}$ is a normalization factor and $\alpha_{i\ell}$ is the complex amplitude associated with the $\ell$-th ray in the $i$-th cluster. The antenna gain at the direction of departure (DoD) azimuth and elevation angles $(\phi_{i\ell}^\mt t,\theta_{i\ell}^\mt t)$, and the direction of arrival (DoA)$ (\phi_{i\ell}^\mt r,\theta_{i\ell}^\mt r)$, are denoted by $\Lambda_\mt r(\phi_{i\ell}^\mt r,\theta_{i\ell}^\mt r)$, and $\Lambda_\mt t(\phi_{i\ell}^\mt r,\theta_{i\ell}^\mt r)$ respectively. The DoDs and DoAs of the scatterers are assumed randomly distributed with a Laplacian distribution~\cite{Ayach2014}. The vectors, $\bd a_\mt r(\phi_{i\ell}^\mt r,\theta_{i\ell}^\mt r)$ and $\bd a_\mt t(\phi_{i\ell}^\mt r,\theta_{i\ell}^\mt r)$ are respectively the receive and transmit array steering vectors associated with the $\ell$-th ray in the $i$-th cluster. For an uniform planar array (UPA) located in the $yz$-plane, the array response is
\begin{align}
    \nonumber\bd a(\phi,\theta)=& \frac{1}{\sqrt{N}}\big[e^{jkd\left( m\sin\left(\phi\right)\sin\left(\theta\right)+n\cos\left(\theta\right)\right)}\big]\\
    & 0 \geq m \geq N_y, 0 \geq n \geq N_z
\end{align}
where $N$ is the total number of elements, while $N_y$ and $N_z$ are the number of grid points in the $y$, and $z$ planes respectively such that $N=N_yN_z$. 

Let the transmit power be divided equally among all the data streams. Then, the mutual information is expressed as
\begin{align}\label{eq:mi_comp}
    \nonumber \mathcal I=\log_2 \Bigg( \Big| \bd I_{N_\mt s}+\frac{\rho}{N_\mt s} \bd R_n^{-1} &\bd W_\mt{BB}^H\bd W_\mt{RF}^H\bd H\bd F_\mt{RF}\bd F_\mt{BB}\\
    &\times \bd F_\mt{BB}^H\bd F_\mt{RF}^H \bd H^H  \bd W_\mt{RF}\bd W_\mt{BB} \Big| \Bigg).
\end{align}
Here, $\bd R_n$ is the noise covariance matrix at the receiver given by \mbox{$\bd R_n=\sigma^2\bd W_\mt{BB}^H \bd W_\mt{RF}^H\bd W_\mt{RF}\bd W_\mt{BB}$}. The optimum beamformer is composed of the precoding and combining matrices ($\bd F_\mt{BB},\bd F_\mt{RF},\bd W_\mt{BB},\bd W_\mt{RF}$) that maximize the mutual information. However, this design problem is a joint non-convex optimization that is intractable. To overcome this difficulty, we decompose it into separate transmit and receive subproblems~\cite{Ayach2014}, which yields the mutual information at the transmit-side
\begin{align}\label{eq:precoding_formulation}
    \mathcal I=\log_2 \left( \left| \bd I_{N_\mt s}+\frac{\rho}{N_\mt s}\bd H\bd F_\mt{RF}\bd F_\mt{BB} \bd F_\mt{BB}^H\bd F_\mt{RF}^H \bd H^H\right| \right).  
\end{align}
Furthermore, by defining a virtual channel for the receive section as $\bd H_\mt r=\bd H\bd F_{\mt{opt}}$ with $\bd F_{\mt{opt}}$ being the optimum precoder employed at the transmit side, the mutual information at the receive-side can be specified as
\begin{align}
    \nonumber\mathcal I(\bd H_\mt r,\bd W_\mt{BB},&\bd W_\mt{RF})=
    \log_2 \Bigg( \Big| \bd I_{N_\mt s}+\frac{\rho}{N_\mt s \sigma^2} \bd H_\mt r^H \bd W_\mt{RF}\bd W_\mt{BB}\\
    &\times (\bd W_\mt{RF}\bd W_\mt{BB} \bd W_\mt{BB}^H\bd W_\mt{RF}^H)^{-1} \bd W_\mt{BB}^H\bd W_\mt{RF}^H \bd H_\mt r\Big | \Bigg).  
\end{align}
Hence, the problem is transformed to enable the separate design of precoding and combining matrices.

In general, the analog precoder and combining matrices, $\bd F_\mt{RF}$ and $\bd W_{\mt{RF}}$, are implemented either using analog phase shifters or analog switches along with RF combiners/splitters. In this paper, we focus on hybrid architectures based on switch networks and consider only the transmit-side.

\section{Switch-based Hybrid Precoder Design }
\label{sec:proposal}
\begin{figure}[!t]
    \centering
    \includegraphics[width=8cm]{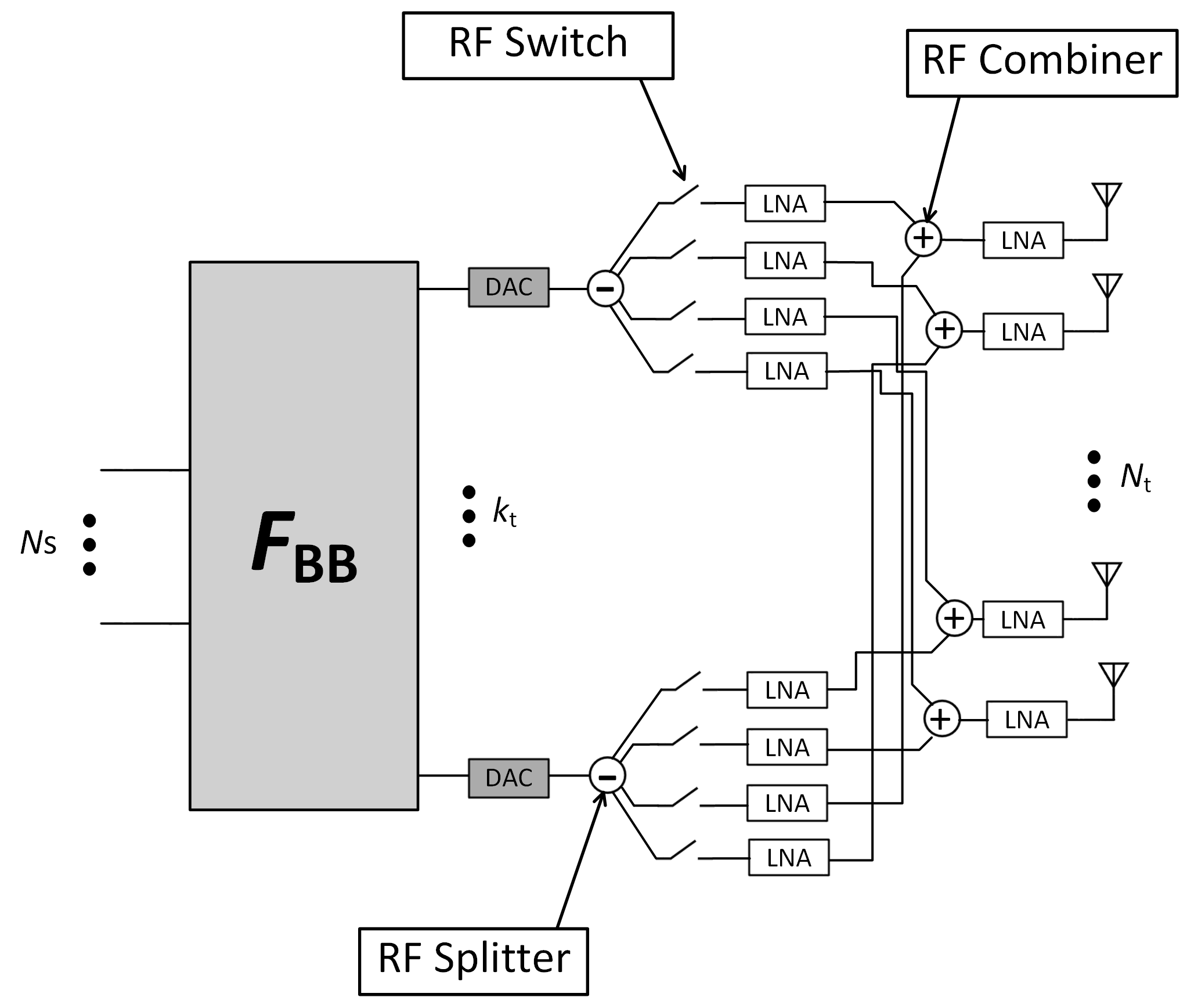}
    \caption{Simplified analog architecture for Hybrid MIMO beamforming with analog switches, combiners, and splitters.}
    \label{fig:combining_diagram}
\end{figure}

Given a hybrid precoder based on a network of analog switches, we sketch the general model for splitting, switching, and combining in Fig.~\ref{fig:combining_diagram}. The $N_\mt s$ data streams are first digitally precoded by $\bd F_\mt{BB}$. Then each of ${k_\mt t}$ precoded signals is converted into the RF domain through a DAC. The RF signal is then split into $N_\mt t$ branches, and each split signal is directed to a low-noise amplifier (LNA) via an RF switch. Finally, in each transmit antenna, a set of ${k_\mt t}$ signals are combined, amplified and transmitted. 

We begin the design by finding the optimum precoding matrix. Given that the channel has a singular value decomposition, e.g., ${\bd H}=\bd U \bs \Sigma \bd V^H$ such that $\bd U$ is an $N_\mt r \times \mt{rank}({\bd H})$ unitary matrix, $\bs \Sigma$ is a $\mt{rank}({\bd H})\times \mt{rank}({\bd H})$ diagonal matrix of descending singular values, and is a unitary matrix. The unconstrained optimum precoder is given by the first $N_\mt s$ singular vectors and the diagonal matrix $\Gamma$ as $\bd F_{\mt{opt}}=\bd V_{N_\mt s}\bs \Gamma$. The diagonal matrix $\Gamma$ is effectively obtained via a water-filling power allocation method. Hence, it is reasonable to split the optimum precoder, consider an equal power allocation ($\bs \Gamma=\bd I$), and apply the power allocation matrix when required. By employing the unconstrained optimum precoder, we have
\begin{align}
    \nonumber\mathcal I({\bd H},\bd F_\mt{opt})=
    \log_2& \Bigg( \Big| \bd I+\frac{\rho}{N_\mt s \sigma^2} {\bd H} \bd F_\mt{opt}\bd F_\mt{opt}^H {\bd H}^H\Big | \Bigg). 
\end{align}
We assume that the mmWave system and propagation channel parameters are selected such that a hybrid precoder $\bd F_\mt{RF}\bd F_\mt{BB}$, sufficiently close to $\bd F_{\mt{opt}}=\bd V_{N_\mt s}$, is attainable. Thus, it is assumed that the matrices $\bd I_{N_\mt s}-\bd V_{N_\mt s}^H\bd F_\mt{RF}\bd F_\mt{BB}\bd F_\mt{BB}^H\bd F_\mt{RF}\bd V_{N_\mt s}$, and $\bd V_{\bar{N}_\mt s}^H\bd F_\mt{RF}\bd F_\mt{BB}$ have a set of sufficiently small eigenvalues~\cite{Ayach2014}. Note that $\bd V_{\bar{N}_\mt s}$ denotes the eigenvectors associated with the subspace complementary to $\bd V_{{N}_\mt s}$. Now by employing Sylvester's determinant theorem, and Schur's complement identity for matrix determinants, we can specify the mutual information as
\begin{align}
    &\mathcal I({\bd H})=
   \nonumber \log_2 \left( \left| \bd I+\frac{\rho}{N_\mt s \sigma^2} \bs \Sigma^2 \bd V^H\bd F_\mt{RF}\bd F_\mt{BB}\bd F_\mt{BB}^H\bd F_\mt{RF}\bd V\right| \right),\\
 \nonumber  &\approx
    \log_2 \left( \left| \bd I+
    \begin{bmatrix}
    \frac{\rho}{N_\mt s \sigma^2}{\bs \Sigma}_{N_\mt s}^2 \bd V_{N_\mt s}^H\bd F_\mt{RF}\bd F_\mt{BB}\bd F_\mt{BB}^H\bd F_\mt{RF}\bd V_{N_\mt s} & 0\\
    0&  0
    \end{bmatrix}
    \right| \right),\\
     \nonumber &=
    \log_2 \left( \left| \bd I_{N_\mt s}+\frac{\rho}{N_\mt s \sigma^2}{\bs \Sigma}_{N_\mt s}^2 \bd V_\mt s^H\bd F_\mt{RF}\bd F_\mt{BB}\bd F_\mt{BB}^H\bd F_\mt{RF}\bd V_\mt s  \right| \right)
    \\
     &=
    \log_2 \left( \left| \bd I_{N_\mt s}+\frac{\rho}{N_\mt s \sigma^2} \bd H_1 \bd F_\mt{RF}\bd F_\mt{BB}\bd F_\mt{BB}^H\bd F_\mt{RF}^H \bd H_1^H    \right| \right)\label{eq:mu_inf_appr_1},
\end{align}
where $\bd H_1$ is the channel constructed by the first $N_\mt s$ singular vectors and singular values of ${\bd H}$. 

If we define a new virtual matrix $\tilde{\bd H}= \bd H_1 \bd F_\mt{RF}$ of size $N_\mt r \times {k_\mt t} $ and singular value decomposition of $\tilde{\bd H}=\tilde{\bd U}\tilde{\bs \Sigma}\tilde{\bd V}^H$, we can maximize (\ref{eq:mu_inf_appr_1}) by having $\bd F_\mt{BB}=\tilde{\bd V}_{N_\mt{s}}$.
\begin{align}
    \nonumber\mathcal I(\tilde{\bd H})&=
    \log_2 \left( \left| \bd I+\frac{\rho}{N_\mt s \sigma^2} \tilde{\bd H}\bd F_\mt{BB}\bd F_\mt{BB}^H \tilde{\bd H}^H    \right| \right)\\
    \nonumber&=\log_2 \left( \left| \bd I+\frac{\rho}{N_\mt s \sigma^2} \tilde{\bd H}\tilde{\bd V}_{N_\mt{s}}\tilde{\bd V}_{N_\mt{s}}^H \tilde{\bd H}^H    \right| \right)\\
    \nonumber&=
    \log_2 \left( \left| \bd I+
    \begin{bmatrix}
    \frac{\rho}{N_\mt s \sigma^2}\tilde{\bs \Sigma}_{N_\mt s}^2 \tilde{\bd V}_{N_\mt s}^H\tilde{\bd V}_{N_\mt s}\tilde{\bd V}_{N_\mt s}^H\tilde{\bd V}_{N_\mt s} & 0\\
    0&  0
    \end{bmatrix}
    \right| \right),\\
   \nonumber &=
  \log_2 \left( \left| \bd I_{N_\mt s}+\frac{\rho}{N_\mt s \sigma^2}\tilde{\bs \Sigma}_{N_\mt s}^2   \right| \right)\\
      \label{eq:log_sv_H_tilde}&=
    \log_2 \left( \left| \bd I_{N_\mt s}+\frac{\rho}{N_\mt s \sigma^2} \tilde{\bd H}_1 \tilde{\bd H}_1^H    \right| \right),
\end{align}
where $\tilde{\bd H}_1$ denotes the new virtual channel representation achieved by the first $N_\mt s$ eigenvalues, e.g., $\tilde{\bd H}_1=\tilde{\bd U}_{N_\mt{s}}\tilde{\bd S}_{N_\mt{s}}\tilde{\bd V}_{N_\mt{s}}^H$. Note that here $\bd F_\mt{BB}=\tilde{\bd V}_{N_\mt{s}}$ can be exactly achieved as we have only a digital beamformer. Now, if we assume that \mbox{$\mt{rank}(\tilde{\bd H})=N_\mt s$}, then $\bd F_\mt{BB}=\tilde{\bd V}_{N_\mt{s}}=\tilde{\bd V}$ becomes a unitary matrix, and therefore we can write (\ref{eq:log_sv_H_tilde}) as
\begin{align}
   \nonumber \mathcal I({\tilde{\bd H}})&=\log_2 \left( \left| \bd I_{N_\mt s}+\frac{\rho}{N_\mt s \sigma^2} \tilde{\bd H} \tilde{\bd H}^H    \right| \right)\\
    &=\log_2 \left( \left| \bd I_{N_\mt s}+\frac{\rho}{N_\mt s \sigma^2} \bd H_1 \bd F_\mt{RF}\bd F_\mt{RF}^H \bd H_1^H    \right| \right).\label{eq:mu_inf_appr_3}
\end{align}

\section{Hybrid Design with Norm Maximization}\label{sec:norm_max}
Noting that (\ref{eq:mu_inf_appr_3}) and (\ref{eq:mu_inf_appr_1}) are equal given that $\mt{rank}(\tilde{\bd H})=N_\mt s$ and $\bd F_\mt{BB}=\tilde{\bd V}_{N_\mt{s}}$, we can approximate (\ref{eq:mu_inf_appr_3}) as~\cite{Ayach2014}:
\begin{align}
    \mathcal I({\tilde{\bd H}})&\approx\log_2 \left( \left| \bd I_{N_\mt s}+\frac{\rho}{N_\mt s \sigma^2}  \bs \Sigma_1^2    \right| \right)-\left(N_{\mt s}-\|\bd V_1^H\bd F_{\mt{RF}}\bd F_{\mt{BB}}\|_F^2 \right)\label{eq:mu_inf_appr_4}.
\end{align}
Hence, to maximize (\ref{eq:mu_inf_appr_3}), we have to maximize $\|\bd V_1^H\bd F_{\mt{RF}}\bd F_{\mt{BB}}\|_F^2$. Furthermore, the rank of $\bd F_{\mt{RF}}$ should necessarily be greater than or equal to $N_s$ in order to have $\mt{rank}(\bd H_1 \bd F_{\mt{ RF}})=N_\mt s$. Additionally, $\bd V_1^{H}$ is a matrix with $N_s$ orthonormal vectors spanning the space of $\tilde{\bd H}$. Thus, the product $\bd P = \bd V_1^{H}\bd F_\mt{RF}$ must have rank $N_\mt s$ and must span the space of $\tilde{\bd H}$. Note that its columns are not orthonormal, but they form a basis for $\mt{span}(\bd P)$. As this space is also spanned by $\bd F_\mt{BB}$, then $\bd P = \bd V_1^{H}\bd F_\mt{RF}\in \mt{span}(\bd F_\mt{BB})$. Therefore, the projection of $\bd P$ onto $\bd F_\mt{BB}$ does not change its Frobenius norm, giving
\begin{align}\label{eq:norm_correspondence}
  \|\bd V_1^H\bd F_{\mt{RF}}\bd F_{\mt{BB}}\|_F^2=\|\bd V_1^H\bd F_{\mt{RF}}\|_F^2.  
\end{align}

The maximization problem can then be cast as,
\begin{subequations}\label{eq:maximization_primal}
\begin{align+}
\nonumber\max_{\bd F_{\mt{RF}}}\;\; &\|\bd V_1^H\bd F_{\mt{RF}}\|_F^2,\\
\mt{s.t.}\;\;\;& \bd F_{\mt{RF}} \in \left\lbrace 0,1\right\rbrace,\\ 
&\mt{rank}(\bd H_1 \bd F_{\mt{RF}}) = N_{\mt s},\\
\label{eq:maximization_primal_norm}& \|\bd F_{\mt{RF}} \bd F_{\mt{BB}} \|_F^2=N_{\mt s}.
\end{align+}
\end{subequations}

Although we could reduce the problem of joint design of $\bd F_{\mt{RF}}$, and $\bd F_{\mt{BB}}$ to that of only optimising $\bd F_{\mt{RF}}$, the maximization in (\ref{eq:maximization_primal}) is non-convex due to norm maximization, the rank constraint, the binary constraint, the norm equality constraint, and the inherent dependence of $\bd F_{\mt{BB}}$ on $\bd F_{\mt{RF}}$. 

 If we relax the rank constraint to $\mt{rank}(\bd H_1 \bd F_{\mt{RF}}) \leq N_{\mt s}$, noting that $\bd H_1$ is of rank $N_{\mt s}$, the only condition that we need to meet is that $\mt{rank}(\bd F_{\mt{RF}}) \geq N_{\mt s}$. 
There exist some algorithms to constrain the rank, i.e., employing a $\mathrm{Trace}(\cdot)$ function as a surrogate for the rank constraint~\cite{Fazel2004}. However, applying this method requires the introduction of a slack variable that leads to an increase in the dimensionality and, subsequently, a higher computational complexity. Therefore, we lift the rank constraint and devise heuristic approaches to address this restriction.
 
We deal with the norm equality constraint (\ref{eq:maximization_primal_norm}) in two steps. To alleviate the binary property of the problem and the complexity that the norm equality constraint causes in conjunction with the rank constraint, first, we exclude the baseband precoder $\bd F_\mt{BB}$ and meet the constraint later by scaling $\bd F_{\mt{BB}}$.

Furthermore, after relaxing the binary constraint on $\bd F_\mt{RF}$, we can reformulate the problem as
\begin{subequations}\label{eq:rank_trace}
\begin{align}
\max_{\bd F_{\mt{RF}}}\;\; &\|\bd V_1^H\bd F_{\mt{RF}}\|_F^2\\
\mt{s.t.}\;\;\;& 0 \leq \bd F_{\mt{RF}}(i,j)\leq 1, \;\;\;\;  i,j=1,...,N_\mt t({k_\mt t})
\end{align}
\end{subequations}  
Thus far, we have been able to relax the non-convex constraints. However, the maximization of the Frobenius norm function in (\ref{eq:rank_trace}) as a convex function is a non-convex problem. Sequential convex programming (SCP) based on iteratively linearizing the convex function is applied to reformulate the non-convex problem as a series of convex subproblems, each of which can be optimally solved using convex programming~\cite{Fazel2004}. 
We formulate the norm maximization by linearization and use a first-order Taylor expansion as a local approximation. Given $f(\bd F_{\mt{RF}})=\|\bd V_1^H\bd F_{\mt{RF}}\|^2_F$, we express this approximation at point $\ell-1$ as,
\begin{align+}\label{eq:scp_linearization}
\nonumber &  f(\bd F_{\mt{RF}}, \bd F_{\mt{RF}}^{(\ell-1)})\\
 &= f(\bd F_{\mt{RF}}^{(\ell-1)}) + \mt{vec}\left(\nabla f(\bd F_{\mt{RF}}^{(\ell-1)})\right)^T \mt{vec}\left(\bd F_{\mt{RF}}-\bd F_{\mt{RF}}^{(\ell-1)}\right)\nonumber \\
  \nonumber&=\|\bd V_1^H\bd F_{\mt{RF}}^{(\ell-1)}\|_F^2\\
 & +2\mt{vec}\left(\mt{real}\left(\bd V_1\bd V_1^H\right)\bd F_{\mt{RF}}^{(\ell-1)}\right)^T\mt{vec}\left(\bd F_{\mt{RF}}-\bd F_{\mt{RF}}^{(\ell-1)}\right),
\end{align+}
where $\mt{vec}(.)$ denotes vectorization, which stacks the matrix column by column. By linearization we transform the Frobenius norm to an affine form to enable the solving of this maximization by SCP. The convex problem to be solved in the $\ell$-th step can be expressed as: 

\begin{align}
\nonumber\max_{\bd F_{\mt{RF}}}\;\; &f(\bd F_{\mt{RF}}, \bd F_{\mt{RF}}^{(\ell-1)})\\
\label{eq:switching_based}\mt{s.t.}\;\;\;& 0 \leq \bd F_{\mt{RF}}(i,j)\leq 1, \;\;\;\;  i,j=1,...,N_\mt t({k_\mt t}).
\end{align}
After initiation, the SCP procedure approaches a local optimum of (\ref{eq:switching_based}) iteratively. 
\subsection{Direction Adjustment and Rank Control }
We now modify the SCP to improve the optimality of the solution while still meeting the rank constraint. As mentioned earlier, we relax the rank constraint. While this relaxation is helpful to decrease computational complexity, there is no guarantee to achieve $\mt{rank}(\bd F_\mt{RF})\geq N_\mt{s}$ and, subsequently , $\mt{rank}(\bd H_1\bd F_\mt{RF})= N_\mt{s}$. To address this uncertainty, after solving (\ref{eq:switching_based}) at each step, we check the rank requirement, and if met, this confirms that the algorithm is searching in the correct direction and we can then proceed to the next step. If the rank constraint is not met, the algorithm begins searching in other directions until it finds a direction that satisfies the rank requirement.

\begin{figure}[!tb]
    \centering
    \includegraphics[trim={7cm 14cm 7cm 20cm},clip,width=6cm]{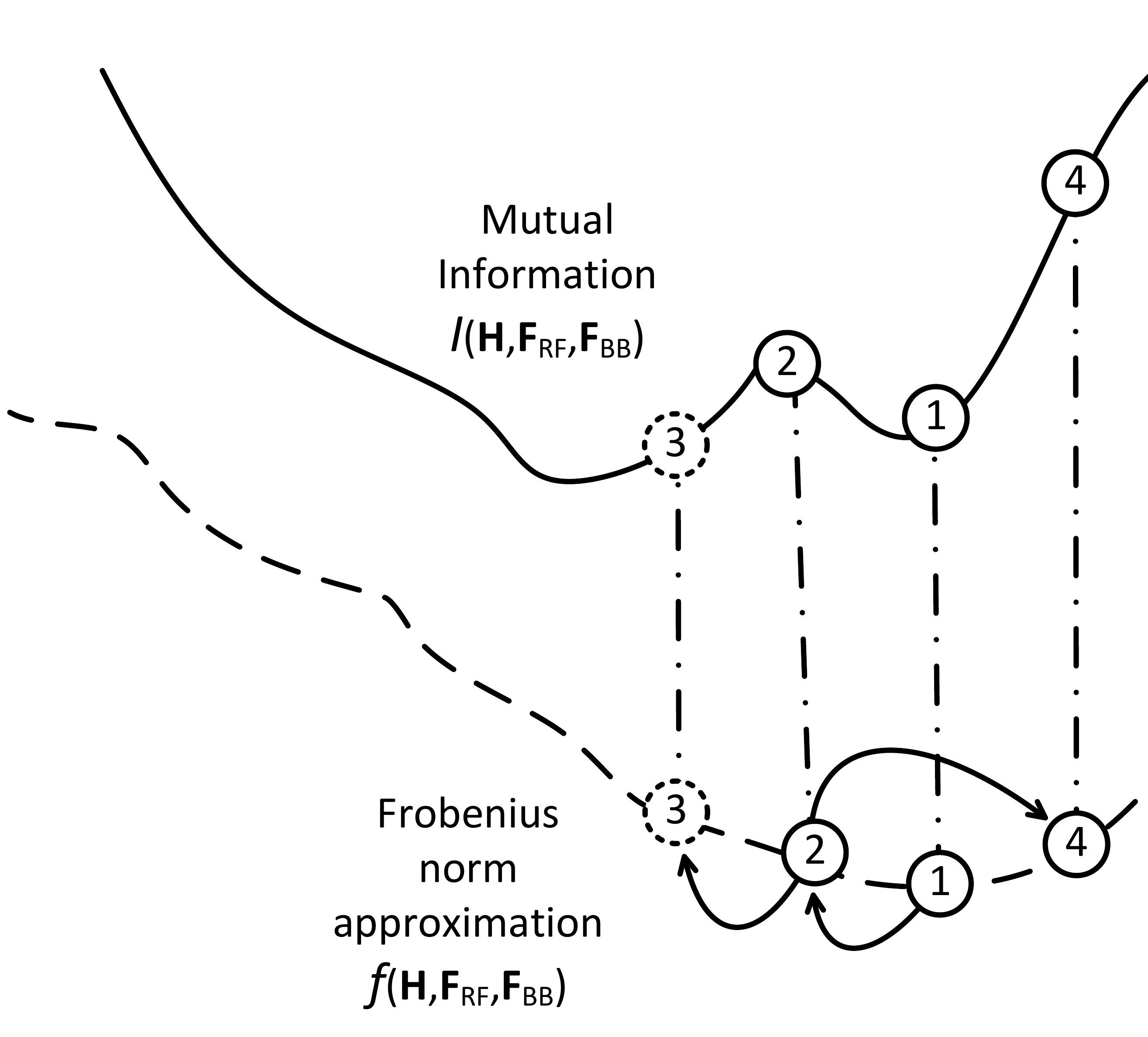}
    \caption{Illustration of the proposed modification to SCP in Algorithm~\ref{alg:SHD-NM}.}
    \label{fig:log_det_norm_approx}
\end{figure}

The ultimate objective of the maximization (\ref{eq:switching_based}) is maximizing the mutual information defined in (\ref{eq:precoding_formulation}). To achieve this goal, we use the Frobenius norm approximation stated in (\ref{eq:mu_inf_appr_4}) under a set of assumptions. These assumptions simply state that the unconstrained precoder $\bd F_\mt{opt}$ is approximately realizable by $\bd F_\mt{RF}\bd F_\mt{BB}$. This assumption is greatly dependent on both the channel and the characteristics of the hybrid precoding network. Such an assumption causes a mismatch between the optimal point given by the Frobenius norm maximization and the actual optimal point. The second modification that we propose addresses this mismatch and leads the SCP algorithm to the point that is a local optimum for both the approximated function and the primal function. We illustrate the concept for this method in Fig.~\ref{fig:log_det_norm_approx}. Given that SCP starts at point 1, the gradient leads the algorithm to point 2, the value of mutual information is assessed at point 2, and since it has been improved at point 2, it is accepted. In the next step, the SCP is lead to point 3. Although the Frobenius norm has increased, the mutual information has decreased. At this point, the algorithm starts searching for other directions to find a point at which both functions are increasing and directs the SCP to point 4. 

The search algorithm that is employed in both modifications above is based on Gaussian randomization. Given that we are searching for a better direction at point $\bd F_\mt{RF}^{(\ell)}$, the algorithm evaluates the direction given by the random variable $\bs{\mathcal{F}}_\mt{RF}$ such that
\begin{align}
    \mt{vec}(\bs{\mathcal{F}}_\mt{RF})\sim \mathcal{N}\left(\mt{vec}\left(\bd F_\mt{RF}^{(\ell)}\right),\bd I_{N_{\mt t}k_{\mt t}}\right) 
\end{align}
Considering the binary characteristics of $\bd F_\mt{RF}$, to find a better direction we search the directions of random variables inside a hypersphere centered at $\bd F_\mt{RF}^{(\ell)}$ with a radius of unity.

\subsection{Baseband Precoder Update}\label{subsec:update}
As explained in ~\ref{sec:norm_max}, we lifted the norm equality constraint, (\ref{eq:maximization_primal_norm}). Now given that $\bd F_\mt{RF}$ is achieved, we have to scale $\bd F_\mt{BB}$ to meet the lifted constraint appropriately. 

To disentangle (\ref{eq:mu_inf_appr_1}) from $\bd F_\mt{BB}$, we assumed that the new virtual channel \mbox{$\tilde{\bd H}_1=\bd H_1\bd F_\mt{RF}$} is of rank $N_\mt s$. Therefore, given the SVD of $\tilde{\bd H}_1$ is $\bd H_1\bd F_\mt{RF}=\tilde{\bd U}\tilde{\bd S}\tilde{\bd V}$, the optimum $\bd F_\mt{BB}$ will be $\bd F_\mt{BB}=\tilde{\bd V}$. Now, having $\bd F_\mt{RF}$ and $\tilde{\bd H}_1$ of a rank $N_\mt s$, we just need to employ SVD to get $\bd F_\mt{BB}$. Next, we apply a scalar adjustment and divide $\bd F_\mt{BB}$ by a scalar value to meet the power constraint as
\begin{align}\label{eq:scalar_adj}
\bd F_\mt{BB}=\frac{\sqrt{N_\mt s} \bd F_\mt{BB}}{\|\bd F_\mt{RF}\bd F_\mt{BB}\|_F}.
\end{align}
Taking $\hat{\bd F}_\mt{RF}$ and $\hat{\bd F}_\mt{BB}$ as the achieved analog precoder and updated baseband precoder, the overall beamformer, $\hat{\bd F}$,  is given by $\hat{\bd F}=\hat{\bd F}_\mt{RF}\hat{\bd F}_\mt{BB}$. It is important to note that $\hat{\bd F}$ results in an equal power allocation if only it is sufficiently close to $\bd F_\mt{opt}$ (see Section~\ref{sec:proposal}). Although the maximization~(\ref{eq:maximization_primal}) minimizes this distance, there is no guarantee that the ultimate distance is sufficiently small. Therefore, the consequent distance between $\hat{\bd F}$, and $\bd F_\mt{opt}$ will result in an unequal power allocation. To address this issue, we combine the decomposition and scaling operations by forming a QR decomposition~\cite{Jiang2018}. Given $\bd F_\mt{RF}$, we assume that we can impose a QR decomposition on $\bd F_\mt{RF}$ as
\begin{align}
    \bd F_\mt{RF}&=\bd U_\mt{RF}\bd R_\mt{RF},
\end{align}
such that
\begin{align}
    \bd U_\mt{RF}&=\bd F_\mt{RF}(\bd F_\mt{RF}^H\bd F_\mt{RF})^{-\frac{1}{2}},\\
       \bd R_\mt{RF}&=(\bd F_\mt{RF}^H\bd F_\mt{RF})^{\frac{1}{2}},
\end{align}
We can then take $\bd F_\mt{BB}$ as
\begin{align}
    \bd F_\mt{BB}&=(\bd F_\mt{RF}^H\bd F_\mt{RF})^{-\frac{1}{2}}\bd G,
\end{align}
and instead find the matrix $\bd G$ that maximizes $\log \left|\bd I+\bd H\bd U_\mt{RF}\bd G\bd G^H\bd U_\mt{RF}^H\bd H^H\right|$. The solution of such a maximization is the first $N_\mt s$ right singular vectors of $\bd H\bd U_\mt{RF}$ that is $\bd G=\bd V^\mt{QR}_\mt{N_\mt s}\bs \Gamma$ such that $\bd H\bd U_\mt{RF}=\bd U^\mt{QR}\bd S^\mt{QR}\bd V^\mt{QR}$.

It is worth noting that $\bd F_\mt{RF}\bd F_\mt{BB}=\bd U_\mt{RF}\bd G$, and since $\bd U_\mt{RF}$ is a semi-unitary matrix ($\bd U_\mt{RF}^H\bd U_\mt{RF}=\bd I$), then $\|\bd F_\mt{RF}\bd F_\mt{BB}\|^2_F=\|\bd U_\mt{RF}\bd G\|^2_F=\|\bd G\|^2_F=N_\mt s$. Hence, the power requirement (norm equality) is already satisfied, and there is no need to apply the scaling. Furthermore, $\bd F_\mt{opt}=\bd U_\mt{RF}\bd G$ is a semi-unitary matrix, and therefore, it allocates the power equally. One caveat is that we have a resultant analog precoder $\bd F_\mt{RF}$   leading to a non-invertible $\bd F_\mt{RF}^H\bd F_\mt{RF}$. We address this by defining a subroutine that checks $\bd F_\mt{RF}^H\bd F_\mt{RF}$ and if it is not invertible, then it is replaced by the closest positive semidefinite matrix.

We list the general form of the proposed modified SCP method in Algorithm~\ref{alg:SHD-NM}. This algorithm takes the channel matrix $\bd H$ and variables $L$, and $I$ as inputs. Then it solves (\ref{eq:switching_based}) for $L$ potential points of increasing values of mutual information, and the Frobenius norm approximation, such that the rank constraint is satisfied. If necessary, the algorithm uses at most $I$ random points to find a better direction. 

\emph{Computational Complexity:} Solving~(\ref{eq:switching_based}) in step~\ref{alg:step_opt} requires $\mathcal{O}(k_\mt t^3N_\mt t^3)$ operations. Moreover, finding the rank of $\tilde{\bd H}$ in step~\ref{alg:step_rank} by SVD needs $\mathcal{O}(k_\mt t N_\mt r^2)$ operations. Also, we need to execute \mbox{$\mathcal{O}(k_\mt t^3+N_\mt s^3)$} operations to calculate the inverse of $(\bd F_\mt{RF}^H\bd F_\mt{RF})$ and SVD of $(\bd U_\mt{RF}\bd U_\mt{RF}^H\bd H\bd H^H\bd U_\mt{RF})$ in order to update $\bd F_\mt{RF}$ in step~\ref{alg:step_update}. Computing~(\ref{eq:precoding_formulation}) in step~\ref{alg:step_mu} involves at least $\mathcal{O}(N_\mt s^3)$ operations. Therefore, the complexity of each iteration of Algorithm~\ref{alg:SHD-NM} is summarized by $\mathcal{O}(k_\mt t^3N_\mt t^3+k_\mt t N_\mt r^2+k_\mt t^3+2N_\mt s^3)$ and can be approximated by $\mathcal{O}(k_\mt t^3N_\mt t^3)$.  

 \begin{algorithm}[!tb]
    \SetKwInOut{Input}{Input}
    \SetKwInOut{Output}{Output}
    \Input{ $\bd H, L, I$}
    Decompose $\bd H=\bd U\bd S\bd V^H$ \\
    Initialize $\bd F_\mt{RF}^0$ at random\\
      \While{$\ell \le L$ and $i \le I$}
    {
      Solve (\ref{eq:switching_based}) and update $\bd F_\mt{RF}^{(\ell)}$ \label{alg:step_opt}\\
       Construct  $\tilde{\bd H}=\bd H_1 \bd F_\mt{RF}^{(\ell)}$\\
       \eIf{$\mt{rank}\left(\tilde{\bd H}\right)=N_\mt s$\label{alg:step_rank}}
              { 
                    \text{Update}  $\bd F_\mt{BB}$\label{alg:step_update}\\
                    Calculate $\mathcal{I}^\ell$ based on (\ref{eq:precoding_formulation})\label{alg:step_mu}
                
                    \eIf{$\mathcal{I}^\ell \geq \mathcal{I}^{\ell-1}$}
                        {
                            $e=\mathcal{I}^\ell-\mathcal{I}^{\ell-1}$\\
                            $\ell=\ell+1$, $i=0$
                        }
                         {
                             \text{Go to step} ~\ref{SIf}
                         }
                }
                        {
              
                    $\text{Update}~\bd F_\mt{RF}^\ell$ by sampling from $\mathcal{N}\left(\mt{vec}\left(\bd F_\mt{RF}^{\ell-1}\right),\bd I\right)$\label{SIf}\\
                    $i=i+1$
            }
    }
    \Output{$\bd F_\mt{RF}, \bd F_\mt{BB}$}

\caption{Switch-based Hybrid Design by Norm Maximization (SHD-NM) } \label{alg:SHD-NM}
\end{algorithm}

\section{Hybrid Design with Majorization}\label{sec:QRQU}
In this section, we propose another method to maximize (\ref{eq:mu_inf_appr_3}) based on majorization theory. 
To begin, we propose a lower bound on (\ref{eq:mu_inf_appr_3}) by the following Theorem.
\begin{theorem}
Assuming that $\tilde{\bd H}$ as a rank deficient matrix can be factorized by a generalized QR decomposition as $\tilde{\bd H}\bd P=\bd Q\bd R$, with $\bd Q,\bd R,\bd P $ being a unitary matrix of size $N_\mt r \times {k_\mt t}$, an upper triangular matrix of size ${k_\mt t} \times {k_\mt t}$, and a permutation matrix of size ${k_\mt t} \times {k_\mt t}$ respectively, then 
\begin{align}
  \mathcal I({\tilde{\bs \Sigma}^2}) \geq  \mathcal I\left(\left|\left[\bd R\right]_{ii}\right|^2\right),
\end{align}
where $\left|[\bd R]_{ii}\right|$ denotes the absolute value of th $i$-th diagonal element of $\bd R$.
\end{theorem}
\begin{IEEEproof}
We know from majorization theory that (see Lemma 4.9 in~\cite{Palomar2006} or~\cite{Jiang2018})
\begin{align}
    \prod_{i=1}^{N\mt s} \Sigma_i^2\geq \prod_{i=1}^{N\mt s} |[\bd R]_{ii}|^2.
\end{align}
Therefore, we can extend this as
\begin{align}
   \nonumber \frac{\rho}{N_\mt s \sigma^2}\prod_{i=1}^{N\mt s}(1+ \Sigma_i^2)&\geq \frac{\rho}{N_\mt s \sigma^2}\prod_{i=1}^{N\mt s} (1+|[\bd R]_{ii}|^2)\\
    \nonumber\log_2\left(\prod_{i=1}^{N\mt s} (1+ \frac{\rho}{N_\mt s \sigma^2}\Sigma_i^2)\right)&\geq \log_2\left(\prod_{i=1}^{N\mt s} (1+\frac{\rho}{N_\mt s \sigma^2}|[\bd R]_{ii}|^2)\right)\\
      \nonumber\mathcal I({\tilde{\bs \Sigma}^2}) &\geq  \mathcal I\left(\left|\left[\bd R\right]_{ii}\right|^2\right)
    \end{align}
\end{IEEEproof}
Now, using the properties of QR decomposition, we can write
\begin{align}\label{eq:quad_diagonal}
  \left|\left[\bd R\right]_{ii}\right|^2= \bd f_{\mt{RF},i}^H\bd A_i \bd f_{\mt{RF},i}
\end{align}
where
\begin{align}
    \nonumber\bd A_i=\bd H_1^H\Pi_{\bd H_1\bd F_\mt{RF}^{i}}\bd H_1,\;\;\Pi_{\bd X}=\bd I-\bd X(\bd X^H\bd X)^{-1}\bd X^H.
\end{align}
The vector $\bd f_{\mt{RF},i}$ denotes the $i$-th column of $\bd F_\mt{RF}$, and the matrix $\bd F_\mt{RF}^{i}$ represents the first $i$ columns of $\bd F_\mt{RF}$.

In summary, we note that the mutual information in (\ref{eq:mu_inf_appr_3}) in terms of $\bd H_1\bd F_\mt{RF}$ is lower bounded by the mutual information given by the diagonal elements of $\bd R$ from a QR decomposition. Thus, we attempt to maximize the diagonal elements of $\bd R$. We cast the maximization for each column of $\bd F_\mt{RF}$ as,

\begin{align+}
\nonumber\max_{\bd f_{\mt{RF},i}}\;\; &\bd f_{\mt{RF},i}^H\bd A_i \bd f_{\mt{RF},i}\\
\label{eq:mu_inf_quad_diag}\mt{s.t.}\;\;\;& 0 \leq \bd f_{\mt{RF},i}(j)\leq 1, \;\;\;\;  j=1,...,N_\mt t
\end{align+}

Using (\ref{eq:mu_inf_quad_diag}), we find $\bd f_{\mt{RF},i}$ that maximizes the diagonal elements of the QR decomposition. As we are just interested in the first $N_\mt s$ diagonal elements, we have a matrix $\bd R$ with $N_\mt s$ nonzero elements on the diagonal, and therefore a rank of $N_\mt s$~\cite{Chan1987}. Due to this property, we assume that the rank constraint is met and remove it from our formulation. We also temporarily remove the transmit power constraint $ \|\bd F_{\mt{RF}} \bd F_{\mt{BB}} \|_F^2=N_{\mt s}$, delivering the condition by updating~$\bd F_{\mt{BB}}$. 

The maximization of the quadratic form in (\ref{eq:mu_inf_quad_diag}) as a convex function is a non-convex problem. 
We formulate the quadratic form maximization by linearization and use a first-order Taylor expansion as a local approximation as in (\ref{eq:scp_linearization}). Given $f(\bd f_{\mt{RF},i})=\bd f_{\mt{RF},i}^H\bd A_i \bd f_{\mt{RF},i}$, we can write this approximation at point $\ell-1$ as,
\begin{align}
  \nonumber f(\bd f_{\mt{RF},i},& \bd f_{\mt{RF},i}^{(\ell-1)})= f(\bd f_{\mt{RF},i}^{(\ell-1)})+\nabla \bd f_{\mt{RF},i}^{(\ell-1)})\left(\bd f_{\mt{RF},i}-\bd f_{\mt{RF},i}^{(\ell-1)}\right)\\
 \nonumber&=f(\bd f_{\mt{RF},i}^{(\ell-1)})+\left((\bd A_i+\bd A_i^T)\bd f_{\mt{RF},i}^{(\ell-1)}\right)\left(\bd f_{\mt{RF},i}-\bd f_{\mt{RF},i}^{(\ell-1)}\right)
\end{align}
 The convex problem to be solved in the $\ell$-th step can be expressed as: 
\begin{align}
\nonumber\max_{\bd f_{\mt{RF},i}}\;\; &f(\bd f_{\mt{RF},i}, \bd f_{\mt{RF},i}^{(\ell-1)})\\
\mt{s.t}\;\;&0\leq\bd f_{\mt{RF},i}\leq 1 \label{eq:switching_based_QR}
\end{align}
We outline the proposed method in Algorithm~\ref{alg:SHD-NM}. We employ the update strategy discussed in subsection ~\ref{subsec:update}.

\emph{Computational Complexity:} Algorithm~\ref{alg:SHD-QRQU} requires $\mathcal{O}(N_\mt t^3)$ operations in each step to solve (\ref{eq:switching_based_QR}) in step~\ref{alg2:step_solve}. Also, updating $\bd A_i$ needs $\mathcal{O}(i^3)$ operations with $i$ denoting the iteration index. Therefore, we summarize the complexity of Algorithm~\ref{alg:SHD-NM} by $\mathcal{O}(k_\mt tN_\mt t^3+\sum\limits_{i=1}^{k_\mt t} i^3)$ and approximate it by $\mathcal{O}(k_\mt tN_\mt t^3)$. 
 \begin{algorithm}[!tb]
    \SetKwInOut{Input}{Input}
    \SetKwInOut{Output}{Output}
    \Input{ $\bd H$}
    Decompose $\bd H=\bd U\bd S\bd V^H$ \\
    Initialize $\Pi_{\bd H_1\bd F_\mt{RF}^{1}}=\bd I$ and $\bd F_\mt{RF}=\bd 0$\\
      \For{$i=1$ to ${k_\mt t}$}
    {
        Initialize $\bd f_{\mt{RF},i}^{(0)}$ at random\\

       \For{$L$ iterations}
    {
      Solve (\ref{eq:switching_based_QR}) and update $\bd f_{\mt{RF},i}^{(\ell)}$ \label{alg2:step_solve} \\
      }
       Update  $[\bd F_\mt{RF}]_i=\bd f_{\mt{RF},i}$, and $\bd A_i$\\
       }
       
       Round $\bd F_\mt{RF}$ and  Construct  $\tilde{\bd H}=\bd F_\mt{RF}\bd H_1$ $\text{and after decomposition  update}\; \bd F_{\mt{BB}}$\label{alg:step_D}\\
    \Output{$\bd F_\mt{RF}, \bd F_{\mt{BB}}^{\mt{D}}, \bd F_{\mt{BB}}^{\mt{LS}}$}

\caption{Switch-based Hybrid Design by QR Decomposition with Quadratic Update (SHD-QRQU) } \label{alg:SHD-QRQU}
\end{algorithm}

\section{Hybrid Precoder Design in Partially Connected Networks}\label{sec:subsets}
By solving (\ref{eq:rank_trace}), we design the analog precoding matrices $\bd F_{\mt{RF}}$, and $\bd F_{\mt{BB}}$ based on a switching network without specific hardware limitations. The proposed formulation allows us to impose arbitrary hardware requirements, a feature that is neither feasible in a greedy method nor viable in dictionary-based techniques.

\begin{figure}[!t]
    \centering
    \includegraphics[width=7 cm]{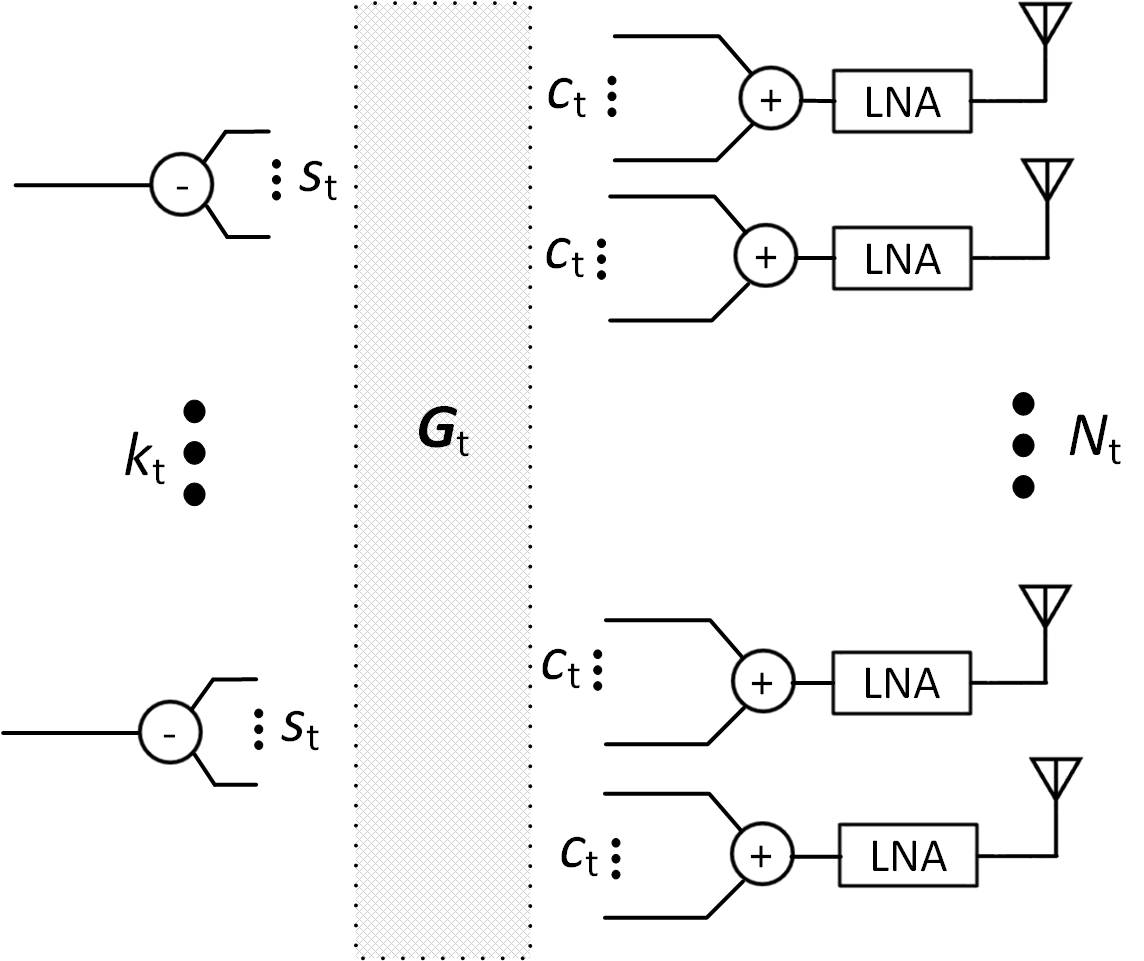}
    \caption{Architecture of a partially connected switch-based analog precoder with $s_\mt t$ outputs for each RF splitter, $c_\mt t$ inputs in each analog combiner and the connectivity matrix $G_\mt t$ that governs the connections.  }
    \label{fig:combining_diagram_Clustering}
\end{figure}

Generally, in a switch-based hybrid precoder and from the structure shown in Fig.~\ref{fig:combining_diagram}, we can define hardware limitations in terms of: (1) The number of outputs of the splitters ($s_\mt t$ in Fig.~\ref{fig:combining_diagram_Clustering}); (2) The number of inputs of the analog precoders in each antenna ($c_\mt t$ in Fig.~\ref{fig:combining_diagram_Clustering}); and (3) The predefined connectivity constraint that allows possible connections between the splitters and combiners ($\bd G_\mt t$ in Fig.~\ref{fig:combining_diagram_Clustering}). 

The mentioned characteristics are inter-related. If we define the set of antennas that are connected to the $\ell$-th splitter via a binary vector $\bd g_{\mt t,\ell}$, we can then concatenate all the ${k_\mt t}$ vectors in the connectivity matrix $ \bd G_\mt t$ as follows,
\begin{align}
    \bd G_\mt t=\left[\bd g_{\mt t,1},\bd g_{\mt t,2},...,\bd g_{\mt t,{k_\mt t}}\right].
\end{align}
Given the connectivity matrix, the other two attributes of the switch network must follow as
\begin{align}
    \bd 1_{N_\mt t}\bd g_{\mt t,\ell}&=s_\mt t,\\
    \bd G_\mt t\bd 1_{{k_\mt t}}&=c_\mt t \bd 1_{N_\mt t}. 
\end{align}
 Furthermore, the total number of connections $n_G$ has to follow
\begin{align}
n_G={k_\mt t} s_\mt t=N_\mt t c_\mt t.
\end{align}

Assuming some hardware constraints are imposed by $\bd G_\mt t, s_\mt t$, and $c_\mt t$, we can introduce a new subspace to optimize (\ref{eq:switching_based}), or (\ref{eq:switching_based_QR}) as
\begin{align}
\bd f_{\mt{RF},i}^T\bar{\bd g}_{\mt t,i}=0,\;\;\;\; i=1,...,{k_\mt t},
\end{align}
where $\bar{\bd g}_{\mt t,i}$ is the Boolean complement of $\bd g_{\mt t,i}$. Therefore, we can cast the problem of switch-based hybrid design in a partially connected network as,
\begin{subequations}
\begin{align}\label{eq:pc}
\nonumber\max_{\bd F_{\mt{RF}}}\;\; &f(\bd F_{\mt{RF}},\bd F_{\mt{RF}}^{(\ell-1)})\\
\mt{s.t.} \;\;\;& 0 \leq \bd F_{\mt{RF}}(i,j)\leq 1, \;\;\;\; \\ 
& \bd f_{\mt{RF},i}^T\bar{\bd g}_{\mt t,i}=0,\;\;\;\; i=1,...,{k_\mt t},
\end{align}
\end{subequations}
 and we then can solve it via Algorithm~\ref{alg:SHD-NM}, or Algorithm~\ref{alg:SHD-QRQU}. 
 
\begin{figure}[!bt]
    \centering
    \includegraphics[width=7 cm]{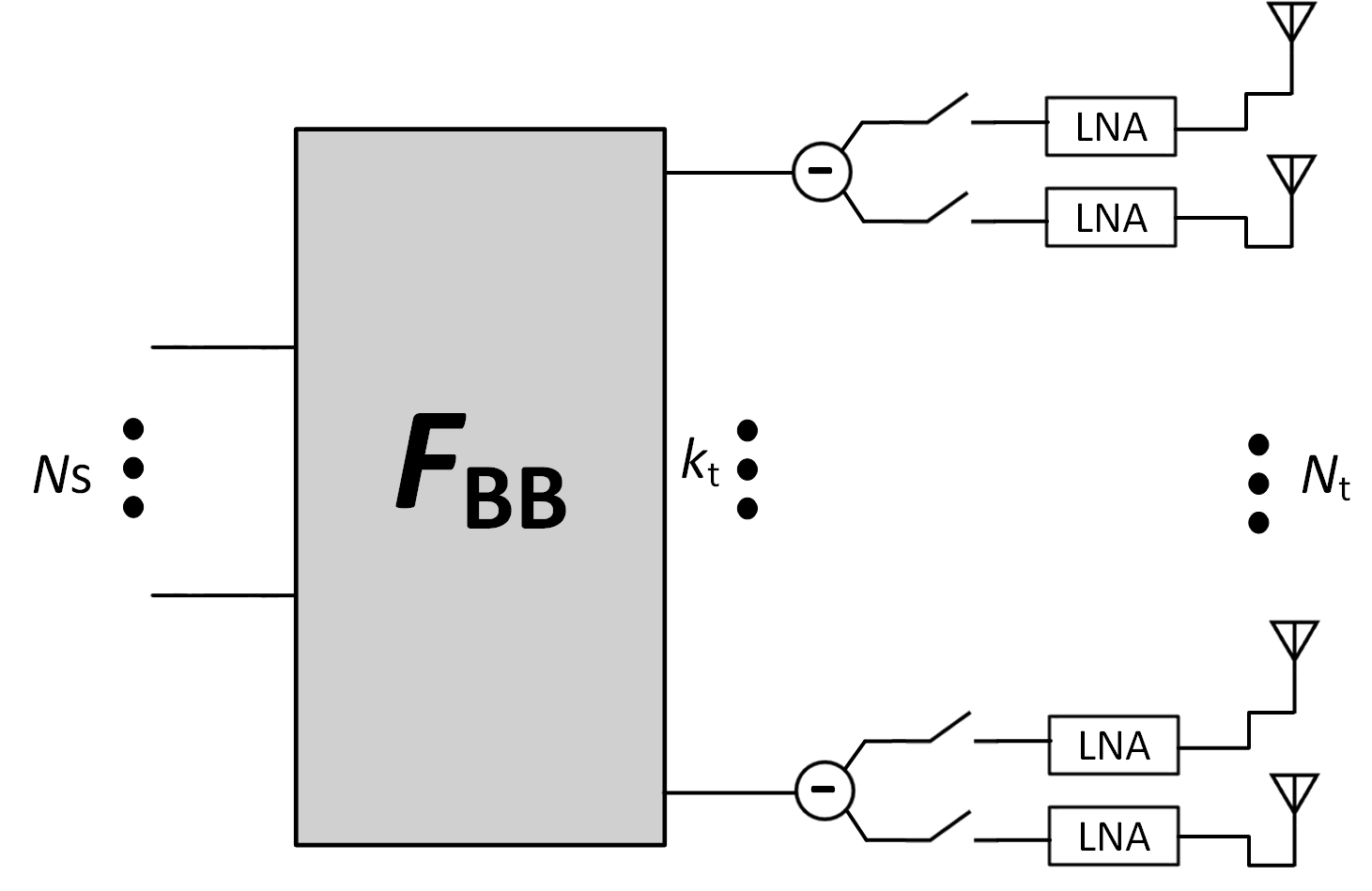}
    \caption{A common structure for partially connected networks: $c_\mt t=1$ with arbitrary $s_\mt t$, and $G_\mt t$.}
    \label{fig:combining_diagram_subset}
\end{figure}

 One common scenario for partially connected switch-based hybrid design is to design for an analog precoder, given that there is no analog combiner in the input of antennas ($c_\mt t=1$) as shown in Fig.~\ref{fig:combining_diagram_subset}.
In this case, $s_\mt t$ is chosen such that ${k_\mt t} s_\mt t=N_\mt t$. Moreover, the connectivity matrix $\bd G_\mt t$ is a set of mutually exclusive columns i.e. $\bd g_{\mt t,i}$. 

\section{Simulation Results}\label{sec:sim}
In this section, we evaluate the performance of the proposed methods through numerical examples. We maximize the spectral efficiency by maximising the mutual information of the transmit side and we assume that there is an ideal combiner at the receiver~(i.e. $\bd W=\bd I$). We use a clustered channel model with $N_\mt{cl}=8$ clusters and $N_\mt{ray}=10$ rays in each cluster with randomly distributed AoDs, and AoAs sampled from a Laplacian distribution~\cite{Xu2002}. We also assume that the complex amplitudes of the rays are sampled from a complex normal distribution with an average power of unity in each cluster. The system model that we adopt is a switch-based hybrid beamformer with $N_\mt t=64$ antennas and ${k_\mt t}=4$ RF chains at the transmit side, and with $N_\mt r=16$ antennas and ${k_\mt t}=4$ RF chains at the receive side. The transmit and receive antenna arrays are uniform planar arrays (UPA) with an inter-element spacing, $d$, of a half-wavelength. We have assumed a sector azimuth angle of $60^\circ$, and sector elevation angle of $30^\circ$ on the transmit side, while on the receive side we assume omni-directional antennas~\cite{Pi2011,Ayach2014}.

We analyze the performance of the described system for different numbers of data streams, $N_\mt s$. For each scenario, we calculate the unconstrained optimal precoder (UOP) achieved by the first $N_\mt s$ eigenmodes of the channel. Moreover, we compare the performance to a phase shifter network by implementing an algorithm called spatially sparse precoding (SSP) proposed in~\cite{Ayach2014} as a fast method to design a hybrid network with phase shifters in the analog section. It is worth noting that the hybrid precoder matrix achieved by SSP does not necessarily provide an equal power allocation for an arbitrary structure. 

In the case of $N_\mt s={k_\mt t}=4$, we implement the greedy algorithm in~\cite{Jiang2018} in switch network mode. We call this algorithm switch-based hybrid design by unified greedy algorithm~(SHD-UG). We then design the switch network with the proposed algorithms, Algorithm~\ref{alg:SHD-NM}, Switch-based Hybrid Design by norm maximization (SHD-NM), and Algorithm~\ref{alg:SHD-QRQU} switch-based hybrid design by QR decomposition with quadratic update (SHD-QRQU). We also implement SHD-NM for designing the switch-based precoder in a partially connected network (SHD-NM-PC). We use the following connectivity matrix for this case,
\begin{align}\label{eq:conn-mat}
    \bd G= &\begin{bmatrix}
1 & 0& 1&0\\
0 & 1& 0&1\\
\vdots & \vdots&\vdots&\vdots\\
1 & 0& 1&0\\
0 & 1& 0&1\\
    \end{bmatrix}.
\end{align}

For convenience of reference, Table~\ref{table:alg_ref} lists the methods that we study by numerical examples. We use $L=1000$, and $I=1000$ when running Algorithm~\ref{alg:SHD-NM}, unless stated otherwise. The spectral efficiency for each value is the averaged value of 100 random channel realizations. Furthermore, we use the CVX package~\cite{grant2008cvx} to solve the convex optimizations in Algorithms~\ref{alg:SHD-NM}, and ~\ref{alg:SHD-QRQU}.

\begin{table}[!b]
\renewcommand{\arraystretch}{1.3}
      \caption{A summary of the methods used in numerical examples.}
      \label{table:alg_ref}
     \centering
\begin{tabular}[!t] {|l|p{5.5cm}| }
 \hline
 Method  & Definition \\ 
  \hline
  UOP & Unconstrained Optimum Precoder  \\ 
  \hline
  SSP & Spatially Sparse Precoder~\cite{Ayach2014}  \\  
  \hline
  SHD-NM &  Switch-based Hybrid Design by Norm Maximization (Algorithm~\ref{alg:SHD-NM})\\ 
    \hline
  SHD-QRQU & Switch-based Hybrid Design by QR Decomposition with Quadratic Update (Algorithm~\ref{alg:SHD-QRQU}) \\ 
 \hline
  SHD-UGD & Switch-based Hybrid Design by Unified Greedy Algorithm~\cite{Jiang2018}  \\ 
 \hline
   SHD-NM-PC & Switch-based Hybrid Design by Norm Maximization in a Partially Connected Network \\ 
 \hline
   SHD-QRQU-PC & Switch-based Hybrid Design by QR Decomposition with Quadratic Update in a Partially Connected Network \\ 
 \hline
\end{tabular}
\end{table}

\begin{figure}[!tb]
    \centering
    \includegraphics[width=9 cm]{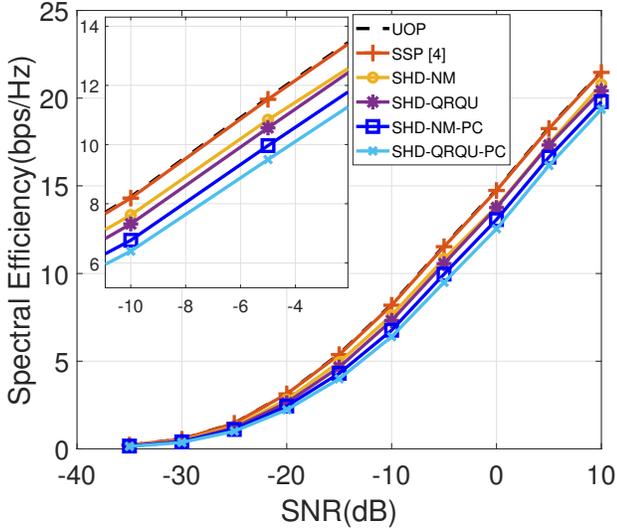}
    \caption{Spectral efficiency achieved by different hybrid design methods for UPAs with $N_\mt t=64$ and $N_\mt r=16$ antennas at transmitter and receiver respectively. The mmWave channel comprises $N_\mt{cl}=8$ clusters and $N_\mt{ray}=10$ rays in each cluster. ${k_\mt t}=4$ RF chains has been used to communicate $N_\mt s=2$ data streams.}
    \label{fig:SE_NS2}
\end{figure}

Fig.~\ref{fig:SE_NS2} shows the spectral efficiency achieved in a $64\times16$ UPA for different values of SNR. Both transmitter sand receivers are assumed to have access to 4 RF chains (${k_\mt t}={k_\mt r}=4$). Also, it is assumed that $N_\mt s=2 $ data streams are transmitted. Fig.~\ref{fig:SE_NS2} illustrates that the proposed method, SHD-NM achieves spectral efficiencies with only a small gap to those achieved by the unconstrained precoder (UOP), and SSP. Considering the significantly lower cost, power, and hardware complexity required by a switch-based hybrid method, such a small gap demonstrates a very good trade-off. Furthermore, the spectral efficiency achieved in a partially connected network sits closely below the fully connected network and introduces yet lower cost, power and complexity. The spectral efficiency attained by SHD-NM for a fully connected network is always superior to that of SHD-QRQU. The same trend holds for partially connected networks as shown by SHD-NM-PC, and SHD-QRQU-PC.

\begin{figure}[!tb]
    \centering
    \includegraphics[width=9 cm]{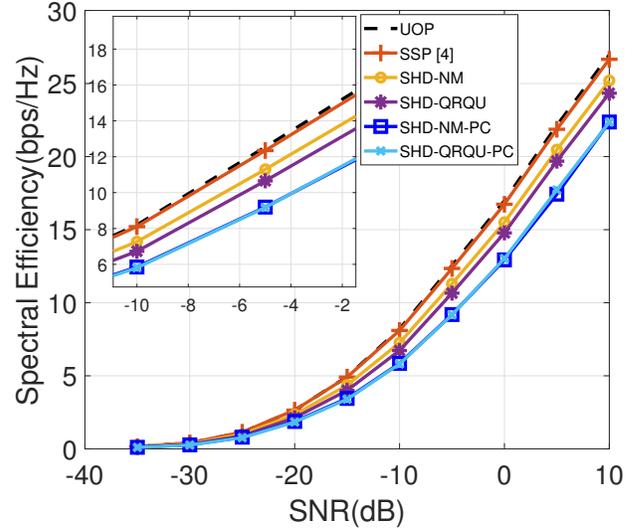}
    \caption{Spectral efficiency achieved by different hybrid design methods for UPAs with $N_\mt t=64$  and $N_\mt r=16$ antennas at transmitter and receiver respectively. The mmWave channel comprises $N_\mt{cl}=8$ clusters and $N_\mt{ray}=10$ rays in each cluster. ${k_\mt t}=4$ RF chains has been used to communicate $N_\mt s=3$ data streams.  }
    \label{fig:SE_NS3}
\end{figure}

In Fig.~\ref{fig:SE_NS3} we study the performance for the case of transmitting $N_\mt s=3 $ data streams. In this scenario, the hybrid network approximates $N_\mt s=3 $ eigenmodes of the channel and the gap between the unconstrained precoder and SSP increases. The switch-based hybrid precoder also experiences larger degradation than that of the phase shift-based strategy, which reflects natural limitations of the switch-based method to approximate 3 eigen-channels with only 4 RF chains and a switch network. Moreover, we can observe that by increasing the number of streams, SHD-QRQU-PC gets closer to SHD-NM-PC.  

\begin{figure}[!tb]
    \centering
    \includegraphics[width=9 cm]{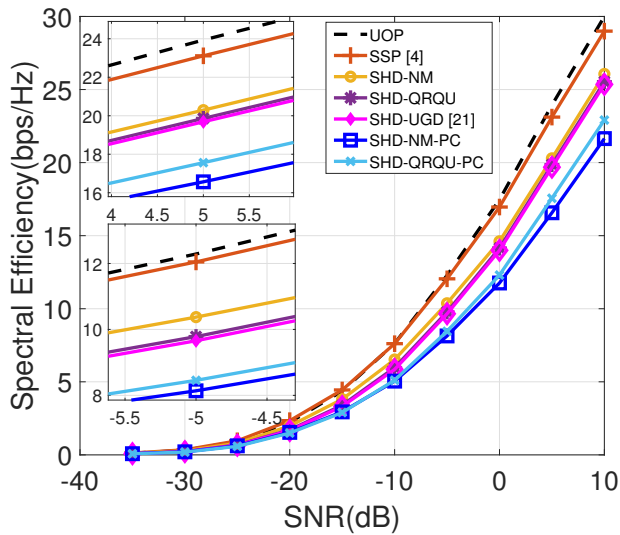}
    \caption{Spectral efficiency achieved by different hybrid design methods for UPAs with $N_\mt t=64$  and $N_\mt r=16$ antennas at transmitter and receiver respectively. The mmWave channel comprises $N_\mt{cl}=8$ clusters and $N_\mt{ray}=10$ rays in each cluster. ${k_\mt t}=4$ RF chains has been used to communicate $N_\mt s=4$ data streams.}
    \label{fig:SE_NS4}
\end{figure}

Next, we examine the performance of the proposed method when $N_\mt s=4$ in Fig.~\ref{fig:SE_NS4}. This scenario is a special case as it is categorized as a hybrid network with ${k_\mt t}=N_\mt s$. This class of hybrid networks has been studied frequently in literature, e.g.~\cite{Sohrabi2016,Jiang2018,Rusu2016}. In this case, the digital precoding matrix $\bd F_\mt{BB}$ becomes a square matrix, and therefore the unitary structure of $\bd F_\mt{BB}$ (as opposed to the generally semi-unitary structure) enables us to decouple the joint design of $\bd F_\mt{RF}, \bd F_\mt{BB}$ and design $\bd F_\mt{RF}$ as an isolated variable. For this scenario, we compare the performance of SHD-NM and SHD-QRQU as a comprehensive solutions with that of the unified greedy algorithm (SHD-UGD) proposed in~\cite{Jiang2018}.

As we expected by increasing $N_\mt s$, the ability of the hybrid network to approximate the optimal unconstrained precoder slightly deteriorates. This can be observed by the increased gap between SSP, and UOP in Fig.~\ref{fig:SE_NS4} when compared to previous cases shown in Figs.~\ref{fig:SE_NS2},~\ref{fig:SE_NS3}. Also, the switch-based network generally  has a larger gap to the unconstrained and phase shift based hybrid structure. As shown in this figure, the proposed algorithms outperform the greedy method (SHD-UGD). While the SHD-NM algorithm provides significantly better performance compared to SHD-UGD, the SHD-QRQU method performs slightly better than SHD-UGD. Another important trend is that SHD-QRQU-PC exhibits better performance compared to that of SHD-NM-PC. This shows that although in a fully-connected network,  SHD-QRQU demonstrates an inferior performance compared to SHD-NM, in a partially-connected network and for certain structures, it can outperform the more computationally expensive SHD-NM. 

\begin{figure}[!tb]
    \centering
    \includegraphics[width=9 cm]{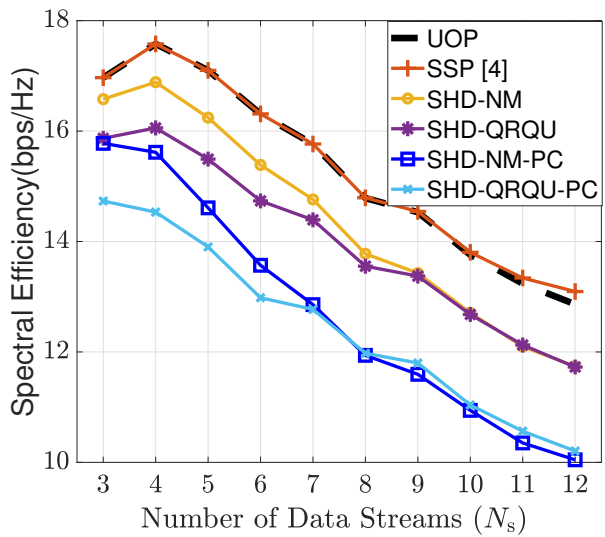}
    \caption{Spectral efficiency achieved by different hybrid design methods for UPAs with $N_\mt t=64$  and $N_\mt r=16$ antennas at transmitter and receiver respectively. The mmWave channel comprises $N_\mt{cl}=8$ clusters and $N_\mt{ray}=10$ rays in each cluster. ${k_\mt t}=12$ RF chains has been used to communicate a varying number of data streams $(N_\mt s=3,...,12)$.}
    \label{fig:SE_NSV}
\end{figure}

We next study the effect of the number of data streams for a fixed value of SNR in Fig.~\ref{fig:SE_NSV}. In this example, we employ a $64\times16$ UPA equipped with 12 RF chains at both transmit and receive sides ($k_\mt t=k_\mt r=12$).  We fix the SNR at 0~dB and run the algorithms for varying values of data streams, i.e. $N_\mt s=3,...,12$. The spectral efficiency decreases with an increasing number of data streams. It is worth noting that the SSP algorithm not only achieves a performance very close to the optimum precoder (UOP), but also it outperforms UOP at higher values of $N_\mt s$. We can account for this by noting that the SSP algorithm does not necessarily provide a precoder with equal power allocation. Hence, SSP can outperform the UOP with an unequal power allocation, which is not desirable. 

The proposed algorithms, SHD-NM and SHD-QRQU, show a reasonable performance by tracking the optimum precoder and maintaining a consistent gap to this optimum. As we observed in Figs.~\ref{fig:SE_NS2},~\ref{fig:SE_NS3},~ and~\ref{fig:SE_NS4}, the SHD-QRQU, gets closer to SHD-NM with increasing $N_\mt s$. It should be noted that we have developed SHD-QRQU based on a QR factorization for a square matrix of an analog precoder. For non-square cases we approximate an invertible $(\bd H_1\bd F_\mt{RF}^i)$ in each step to be able to compute $\bd A_i$. As $N_\mt s$ gets closer to $k_\mt t$, the analog precoder gets closer to a square matrix, and it leads to better performance. Moreover, the feasible solution space shrinks by increasing the number of data streams as the digital beamformer consumes more DoFs provided by the virtual eigen-channels (including the analog precoder and the channel). In this case, the capability of SHD-NM deteriorates since it searches a smaller space for modifying the direction. The boosted performance of SHD-QRQU algorithm for a square analog precoder matrix is confirmed in partially connected networks. The SHD-NM-PC achieves higher spectral efficiency in lower values of $N_\mt s$, i.e. $N_\mt s \leq 8$. However, the spectral efficiency given by SHD-QRQU-PC surpasses that of SHD-NM-PC for higher values of $N_\mt s$.         

\begin{figure}[!tb]
    \centering
    \includegraphics[width=9 cm]{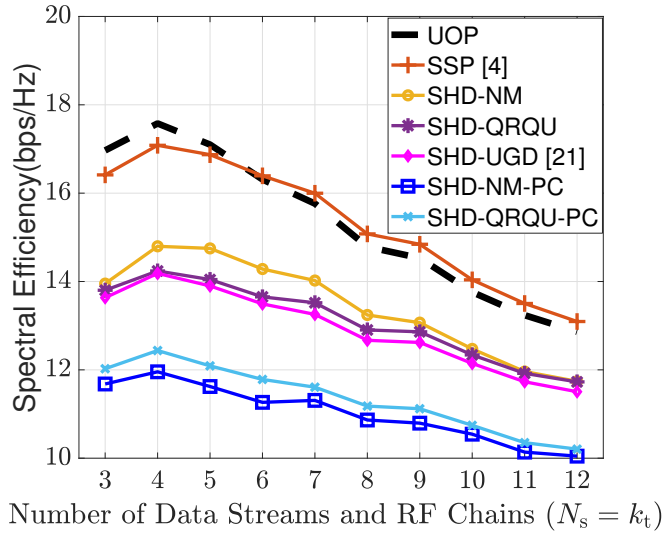}
    \caption{Spectral efficiency achieved by different hybrid design methods for UPAs with $N_\mt t=64$  and $N_\mt r=16$ antennas at transmitter and receiver respectively. The mmWave channel comprises $N_\mt{cl}=8$ clusters and $N_\mt{ray}=10$ rays in each cluster. Varying number of data streams, i.e. $N_\mt s=1,...,12$, are communicated by a similar number of RF chains, $N_\mt s=k_\mt t$ }
    \label{fig:NSKTV}
\end{figure}

Finally, we investigate the behaviours of the algorithms for a scenario of a varying number of streams in a hybrid network with $k_\mt t=N_\mt s$ in Fig.~\ref{fig:NSKTV}. Similar to the previous example, we employ a $64\times16$ UPA and fix the SNR at 0~dB. We then run the algorithms for varying values of $k_\mt t=k_\mt r=N_\mt s=3,...,12$.  
In this example, the number of network structures that lead to an unequal power allocation for SSP algorithm increases significantly. Moreover, the proposed algorithms demonstrate better performance compared to the greedy algorithm (SHD-UGD). The SHD-QRQU algorithm overall performs closer to SHD-NM, and their performance converges with an increasing number of data streams, as explained for the previous example. Moreover, in a partially connected network, SHD-QRQU-PC performs better than SHD-NM as a square analog precoder matrix is always available.         
\section{Conclusion}
\label{sec:conclusion}
We proposed a solution for switch-based hybrid beamforming design for communication in mmWave bands. The binary structure of this type of low-cost, low-power, and low-complexity hybrid design raises new challenges for designing optimal analog, and digital, beamformers. We proposed a new method to decouple the problem of joint optimization of the analog and digital beamformer by confining the problem to a rank-constrained subspace. We proposed two methods to solve the problem effectively. 
 We then introduced linear constraints to include frequently used switch-based structures in partially connected networks. Finally, we examined the effectiveness of the proposed method using a set of numerical examples. The results showed that the proposed methods are feasible, providing an optimal solution for a variety of structures that have an important role as an effective and comprehensive tool in the study of differing relevant scenarios. Furthermore, examining different structures showed that the Switch-based Hybrid Design by Norm Maximization (SHD-NM) algorithm displays a superior performance at a higher complexity compared to Switch-based Hybrid Design by QR Decomposition with Quadratic Update (SHD-QRQU). However, the SHD-QRQU algorithm demonstrates better performance for some structures along with lower complexity.

\section*{Acknowledgement}
This research includes computations using the computational cluster Katana supported by Research Technology Services at UNSW Sydney.

\bibliographystyle{IEEEtran}
\balance
\bibliography{ref.bib}

\begin{thebibliography}{10}
\providecommand{\url}[1]{#1}
\csname url@samestyle\endcsname
\providecommand{\newblock}{\relax}
\providecommand{\bibinfo}[2]{#2}
\providecommand{\BIBentrySTDinterwordspacing}{\spaceskip=0pt\relax}
\providecommand{\BIBentryALTinterwordstretchfactor}{4}
\providecommand{\BIBentryALTinterwordspacing}{\spaceskip=\fontdimen2\font plus
\BIBentryALTinterwordstretchfactor\fontdimen3\font minus
  \fontdimen4\font\relax}
\providecommand{\BIBforeignlanguage}[2]{{%
\expandafter\ifx\csname l@#1\endcsname\relax
\typeout{** WARNING: IEEEtran.bst: No hyphenation pattern has been}%
\typeout{** loaded for the language `#1'. Using the pattern for}%
\typeout{** the default language instead.}%
\else
\language=\csname l@#1\endcsname
\fi
#2}}
\providecommand{\BIBdecl}{\relax}
\BIBdecl

\bibitem{Han2015}
S.~Han, C.~lin I, Z.~Xu, and C.~Rowell, ``Large-scale antenna systems with
  hybrid analog and digital beamforming for millimeter wave 5g,'' \emph{{IEEE}
  Communications Magazine}, vol.~53, no.~1, pp. 186--194, jan 2015.

\bibitem{Zhang2015}
J.~Zhang, X.~Huang, V.~Dyadyuk, and Y.~Guo, ``Massive hybrid antenna array for
  millimeter-wave cellular communications,'' \emph{{IEEE} Wireless
  Communications}, vol.~22, no.~1, pp. 79--87, feb 2015.

\bibitem{Heath2016}
R.~W. Heath, N.~Gonzalez-Prelcic, S.~Rangan, W.~Roh, and A.~M. Sayeed, ``An
  overview of signal processing techniques for millimeter wave {MIMO}
  systems,'' \emph{{IEEE} Journal of Selected Topics in Signal Processing},
  vol.~10, no.~3, pp. 436--453, apr 2016.

\bibitem{Ayach2014}
O.~E. Ayach, S.~Rajagopal, S.~Abu-Surra, Z.~Pi, and R.~W. Heath, ``Spatially
  sparse precoding in millimeter wave {MIMO} systems,'' \emph{{IEEE}
  Transactions on Wireless Communications}, vol.~13, no.~3, pp. 1499--1513, mar
  2014.

\bibitem{Rusu2016}
C.~Rusu, R.~Mendez-Rial, N.~Gonzalez-Prelcic, and R.~W. Heath, ``Low complexity
  hybrid precoding strategies for millimeter wave communication systems,''
  \emph{{IEEE} Transactions on Wireless Communications}, vol.~15, no.~12, pp.
  8380--8393, dec 2016.

\bibitem{Sohrabi2015}
F.~Sohrabi and W.~Yu, ``Hybrid digital and analog beamforming design for
  large-scale {MIMO} systems,'' in \emph{2015 {IEEE} International Conference
  on Acoustics, Speech and Signal Processing ({ICASSP})}.\hskip 1em plus 0.5em
  minus 0.4em\relax {IEEE}, apr 2015.

\bibitem{Bogale2014}
T.~E. Bogale and L.~B. Le, ``Beamforming for multiuser massive {MIMO} systems:
  Digital versus hybrid analog-digital,'' in \emph{2014 {IEEE} Global
  Communications Conference}.\hskip 1em plus 0.5em minus 0.4em\relax {IEEE},
  dec 2014.

\bibitem{Alkhateeb2014}
A.~Alkhateeb, J.~Mo, N.~Gonzalez-Prelcic, and R.~W. Heath, ``{MIMO} precoding
  and combining solutions for millimeter-wave systems,'' \emph{{IEEE}
  Communications Magazine}, vol.~52, no.~12, pp. 122--131, dec 2014.

\bibitem{Nsenga2010}
J.~Nsenga, A.~Bourdoux, and F.~Horlin, ``Mixed analog/digital beamforming for
  60 {GHz} {MIMO} frequency selective channels,'' in \emph{2010 {IEEE}
  International Conference on Communications}.\hskip 1em plus 0.5em minus
  0.4em\relax {IEEE}, may 2010.

\bibitem{Roh2014}
W.~Roh, J.-Y. Seol, J.~Park, B.~Lee, J.~Lee, Y.~Kim, J.~Cho, K.~Cheun, and
  F.~Aryanfar, ``Millimeter-wave beamforming as an enabling technology for 5g
  cellular communications: theoretical feasibility and prototype results,''
  \emph{{IEEE} Communications Magazine}, vol.~52, no.~2, pp. 106--113, feb
  2014.

\bibitem{Gholam2011}
F.~Gholam, J.~Via, and I.~Santamaria, ``Beamforming design for simplified
  analog antenna combining architectures,'' \emph{{IEEE} Transactions on
  Vehicular Technology}, vol.~60, no.~5, pp. 2373--2378, 2011.

\bibitem{Pi2012}
Z.~Pi, ``Optimal transmitter beamforming with per-antenna power constraints,''
  in \emph{2012 {IEEE} International Conference on Communications
  ({ICC})}.\hskip 1em plus 0.5em minus 0.4em\relax {IEEE}, jun 2012.

\bibitem{Zhang2005}
X.~Zhang, A.~Molisch, and S.-Y. Kung, ``Variable-phase-shift-based
  {RF}-baseband codesign for {MIMO} antenna selection,'' \emph{{IEEE}
  Transactions on Signal Processing}, vol.~53, no.~11, pp. 4091--4103, nov
  2005.

\bibitem{Venkateswaran2010}
V.~Venkateswaran and A.-J. van~der Veen, ``Analog beamforming in {MIMO}
  communications with phase shift networks and online channel estimation,''
  \emph{{IEEE} Transactions on Signal Processing}, vol.~58, no.~8, pp.
  4131--4143, aug 2010.

\bibitem{Sayeed2010}
A.~Sayeed and N.~Behdad, ``Continuous aperture phased {MIMO}: Basic theory and
  applications,'' in \emph{2010 48th Annual Allerton Conference on
  Communication, Control, and Computing (Allerton)}.\hskip 1em plus 0.5em minus
  0.4em\relax {IEEE}, sep 2010.

\bibitem{Brady2013}
J.~Brady, N.~Behdad, and A.~M. Sayeed, ``Beamspace {MIMO} for millimeter-wave
  communications: System architecture, modeling, analysis, and measurements,''
  \emph{{IEEE} Transactions on Antennas and Propagation}, vol.~61, no.~7, pp.
  3814--3827, jul 2013.

\bibitem{Sayeed2002}
A.~Sayeed, ``Deconstructing multiantenna fading channels,'' \emph{{IEEE}
  Transactions on Signal Processing}, vol.~50, no.~10, pp. 2563--2579, oct
  2002.

\bibitem{Gao2018}
X.~Gao, L.~Dai, S.~Zhou, A.~M. Sayeed, and L.~Hanzo, ``Beamspace channel
  estimation for wideband millimeter-wave {MIMO} with lens antenna array,'' in
  \emph{2018 {IEEE} International Conference on Communications ({ICC})}.\hskip
  1em plus 0.5em minus 0.4em\relax {IEEE}, may 2018.

\bibitem{Mo2014}
J.~Mo and R.~W. Heath, ``High {SNR} capacity of millimeter wave {MIMO} systems
  with one-bit quantization,'' in \emph{2014 Information Theory and
  Applications Workshop ({ITA})}.\hskip 1em plus 0.5em minus 0.4em\relax
  {IEEE}, feb 2014.

\bibitem{Molisch2017}
A.~F. Molisch, V.~V. Ratnam, S.~Han, Z.~Li, S.~L.~H. Nguyen, L.~Li, and
  K.~Haneda, ``Hybrid beamforming for massive {MIMO}: A survey,'' \emph{{IEEE}
  Communications Magazine}, vol.~55, no.~9, pp. 134--141, 2017.

\bibitem{Jiang2018}
Y.~Jiang, Y.~Feng, and M.~K. Varanasi, ``Hybrid beamforming for massive {MIMO}:
  A unified solution for both phase shifter and switch networks,'' in
  \emph{2018 10th International Conference on Wireless Communications and
  Signal Processing ({WCSP})}.\hskip 1em plus 0.5em minus 0.4em\relax {IEEE},
  oct 2018.

\bibitem{Mendez-Rial2016}
R.~Mendez-Rial, C.~Rusu, N.~Gonzalez-Prelcic, A.~Alkhateeb, and R.~W. Heath,
  ``Hybrid {MIMO} architectures for millimeter wave communications: Phase
  shifters or switches?'' \emph{{IEEE} Access}, vol.~4, pp. 247--267, 2016.

\bibitem{Poon2012}
A.~S.~Y. Poon and M.~Taghivand, ``Supporting and enabling circuits for antenna
  arrays in wireless communications,'' \emph{Proceedings of the {IEEE}}, vol.
  100, no.~7, pp. 2207--2218, jul 2012.

\bibitem{Ahmed2018}
I.~Ahmed, H.~Khammari, A.~Shahid, A.~Musa, K.~S. Kim, E.~D. Poorter, and
  I.~Moerman, ``A survey on hybrid beamforming techniques in 5g: Architecture
  and system model perspectives,'' \emph{{IEEE} Communications Surveys {\&}
  Tutorials}, vol.~20, no.~4, pp. 3060--3097, 2018.

\bibitem{Molisch2004}
A.~Molisch and M.~Win, ``{MIMO} systems with antenna selection,'' \emph{{IEEE}
  Microwave Magazine}, vol.~5, no.~1, pp. 46--56, mar 2004.

\bibitem{Gharavi-Alkhansari2004}
M.~Gharavi-Alkhansari and A.~Gershman, ``Fast antenna subset selection in
  {MIMO} systems,'' \emph{{IEEE} Transactions on Signal Processing}, vol.~52,
  no.~2, pp. 339--347, feb 2004.

\bibitem{Nosrati2017}
H.~Nosrati, E.~Aboutanios, and D.~B. Smith, ``Receiver-transmitter pair
  selection in {MIMO} phased array radar,'' in \emph{2017 {IEEE} International
  Conference on Acoustics, Speech and Signal Processing ({ICASSP})}.\hskip 1em
  plus 0.5em minus 0.4em\relax {IEEE}, mar 2017.

\bibitem{Wang2014}
X.~Wang, E.~Aboutanios, M.~Trinkle, and M.~G. Amin, ``Reconfigurable adaptive
  array beamforming by antenna selection,'' \emph{{IEEE} Transactions on Signal
  Processing}, vol.~62, no.~9, pp. 2385--2396, may 2014.

\bibitem{Nosrati2017a}
H.~Nosrati, E.~Aboutanios, and D.~B. Smith, ``Array spatial thinning for
  interference mitigation by semidefinite programming,'' in \emph{2017 25th
  European Signal Processing Conference ({EUSIPCO})}.\hskip 1em plus 0.5em
  minus 0.4em\relax {IEEE}, aug 2017.

\bibitem{Amin2016}
M.~G. Amin, X.~Wang, Y.~D. Zhang, F.~Ahmad, and E.~Aboutanios, ``Sparse arrays
  and sampling for interference mitigation and {DOA} estimation in {GNSS},''
  \emph{Proceedings of the {IEEE}}, vol. 104, no.~6, pp. 1302--1317, jun 2016.

\bibitem{Ayach2013}
O.~E. Ayach, R.~W. Heath, S.~Rajagopal, and Z.~Pi, ``Multimode precoding in
  millimeter wave {MIMO} transmitters with multiple antenna sub-arrays,'' in
  \emph{2013 {IEEE} Global Communications Conference ({GLOBECOM})}.\hskip 1em
  plus 0.5em minus 0.4em\relax {IEEE}, dec 2013.

\bibitem{Nosrati2018}
H.~Nosrati, E.~Aboutanios, and D.~B. Smith, ``Spatial array thinning for
  interference cancellation under connectivity constraints,'' in \emph{2018
  {IEEE} International Conference on Acoustics, Speech and Signal Processing
  ({ICASSP})}.\hskip 1em plus 0.5em minus 0.4em\relax {IEEE}, apr 2018.

\bibitem{Mendez-Rial2015}
R.~Mendez-Rial, C.~Rusu, A.~Alkhateeb, N.~Gonzalez-Prelcic, and R.~W. Heath,
  ``Channel estimation and hybrid combining for {mmWave}: Phase shifters or
  switches?'' in \emph{2015 Information Theory and Applications Workshop
  ({ITA})}.\hskip 1em plus 0.5em minus 0.4em\relax {IEEE}, feb 2015.

\bibitem{Fazel2004}
M.~Fazel, H.~Hindi, and S.~Boyd, ``Rank minimization and applications in system
  theory,'' in \emph{Proceedings of the 2004 American Control
  Conference}.\hskip 1em plus 0.5em minus 0.4em\relax {IEEE}, 2004.

\bibitem{Palomar2006}
D.~P. Palomar and Y.~Jiang, ``{MIMO} transceiver design via majorization
  theory,'' \emph{Foundations and Trends{\textregistered} in Communications and
  Information Theory}, vol.~3, no. 4-5, pp. 331--551, 2006.

\bibitem{Chan1987}
T.~F. Chan, ``Rank revealing {QR} factorizations,'' \emph{Linear Algebra and
  its Applications}, vol. 88-89, pp. 67--82, apr 1987.

\bibitem{Xu2002}
H.~Xu, V.~Kukshya, and T.~Rappaport, ``Spatial and temporal characteristics of
  60-{GHz} indoor channels,'' \emph{{IEEE} Journal on Selected Areas in
  Communications}, vol.~20, no.~3, pp. 620--630, apr 2002.

\bibitem{Pi2011}
Z.~Pi and F.~Khan, ``An introduction to millimeter-wave mobile broadband
  systems,'' \emph{{IEEE} Communications Magazine}, vol.~49, no.~6, pp.
  101--107, jun 2011.

\bibitem{grant2008cvx}
M.~Grant, S.~Boyd, and Y.~Ye, ``{CVX}: Matlab software for disciplined convex
  programming,'' 2008.

\bibitem{Sohrabi2016}
F.~Sohrabi and W.~Yu, ``Hybrid digital and analog beamforming design for
  large-scale antenna arrays,'' \emph{{IEEE} Journal of Selected Topics in
  Signal Processing}, vol.~10, no.~3, pp. 501--513, apr 2016.

\end{thebibliography}

\end{document}